\documentstyle[11pt,epsfig]{article}
\def\be{\begin{eqnarray}}
\def\ee{\end{eqnarray}}
\def\MeV{\mbox{MeV}}

\def\roughly#1{\mathrel{\raise.3ex\hbox{$#1$\kern-.75em%
\lower1ex\hbox{$\sim$}}}}
\def\lsim{\roughly<}
\def\gsim{\roughly>}
\def\la{\langle}
\def\ra{\rangle}
\def\Tr{\rm Tr}
\newcommand{\Slash}[1]{\ooalign{\hfil/\hfil\crcr$#1$}}
\def\del{\partial}

\def\bi{\bibitem}

\begin{document}
\centerline{\large\bf Flavor Symmetry and Topology Change}

\centerline{\large\bf in Nuclear Symmetry Energy for Compact Stars}

\vskip 1cm
\vspace{.30cm}
\begin{center}
Hyun Kyu Lee$^{*, \ddag}$ and Mannque Rho$^{*,\dagger}$

\vskip 0.50cm
\noindent { \it $^*$Department of Physics, Hanyang University, Seoul 133-791, Korea}

\noindent{ \it $^{\ddag}$ Asia Pacific Center for Theoretical Physics, 790-784 Pohang, Korea}

\noindent { \it $^{\dagger}$Institut de Physique Th$\acute{e}$orique, CEA Saclay, 91191 Gif-sur-Yvette, France}

\end{center}
\vskip 1cm
\centerline{\bf ABSTRACT}
\vskip 0.5cm
\noindent The nuclear symmetry energy figures crucially in the structure of asymmetric nuclei and, more importantly, in the equation of state (EoS) of compact stars. At present it is almost totally unknown, both experimentally and theoretically, in the density regime appropriate for the interior of neutron stars. Basing on a strong-coupled structure of dense baryonic matter encoded in the skyrmion crystal approach with a topology change and resorting to the notion of generalized HLS (hidden local symmetry) in hadronic interactions, we address a variety of hitherto unexplored issues of nuclear interactions associated with the symmetry energy, i.e., kaon condensation and hyperons, possible topology change in dense matter, nuclear tensor forces, conformal symmetry and chiral symmetry etc in the EoS of dense compact-star matter. One of the surprising results coming from the hidden local symmetry structure that is distinct from what is given by standard phenomenological approaches is that at high density,  baryonic matter is driven by RG flow to the ``dilaton-limit fixed point (DLFP)" constrained by ``mended symmetries." We further propose how to formulate kaon condensation and hyperons in compact-star matter in a framework anchored on a single effective Lagrangian by treating hyperons as the Callan-Klebanov kaon-skyrmion bound states simulated on crystal lattice. This formulation suggests that hyperons can figure in the stellar matter -- if at all --  {\em when or after} kaons condense, in contrast to the standard phenomenological approaches where the hyperons appear as the first strangeness degree of freedom in matter thereby suppressing or delaying kaon condensation.  In our simplified description of the stellar structure in terms of symmetry energies which is compatible with that of the 1.97 solar mass star, kaon condensation plays a role of ``doorway state" to strange-quark matter.
\newpage
\tableofcontents
\newpage

\section{Introduction}
\indent\indent The question as to whether strong interactions have anything to do with black hole formation or rather with the other side of the coin, the maximum mass of compact stars stable against collapse to black holes, can be addressed in terms of the nuclear symmetry energy $\epsilon_{sym}$. How nuclear forces mediating the strong interactions and controlling the phase structure of multi-baryonic systems enter into the nature of $\epsilon_{sym}$, presumably controlling the fate of compacts stars, is one of the principal themes of our current theoretical research into baryonic matter at high density. This note is a sequel to, and a substantial updating of, the previous reports~\cite{WCU1,WCU2}.

On the experimental side, there are broadly two avenues, namely, terrestrial laboratories and space observatories. The heavy-ion collision is one of the terrestrial tools to gain information on how the strong interactions between hadrons (nuclear interaction) are modified with varying densities. In project are such laboratories as  KoRIA (Korean Rare Isotope Accelerator, officially called ``RAON") and other RIB machines, FAIR (Facility for Antiproton and Ion Reactions) at GSI/Darmstadt, FRIB(Facility for Rare Isotope Beams) at MSU/Michigan, and NICA (Nuclotron-based Ion Collider Facility) at JINR/Dubna.  They will probe the ranges of temperature up to $50-60$ MeV and of density up to $(3-4) n_0$ (where $n_0$ is the normal nuclear matter density $n_0\approx 0.16$ fm$^{-3}$) in the nucleon number density\cite{CHP}\footnote{In this paper, unless otherwise noted, baryon number density will be denoted $n$.} commensurate with the temperature and density encountered in binary mergers of neutron stars and black holes~\cite{sekiguchi} within the transient time of $10-20$ fm/sec~\cite{toro}.  The $n$-$p$ asymmetry, $\delta ( =\frac{N-P}{N+P}= 1-2x, x=P/(N+P))$, ranges from $0$ for $^{12}$C  to $0.198$ for $^{197}$Au or 0.227 for $^{238}U$ collisions~\cite{CPM}.

In heavy-ion collisions the transient time for the dense matter phase is too short to activate the weak interactions. Hence the initial $n$-$p$ symmetry factor $\delta$ does not change, so any process involving net strangeness number production is suppressed.
This difference between the hadronic matter in heavy ion collision and the hadronic matter in weak equilibrium of stellar matter renders the role of the symmetry energy different from each other.

For the stellar mass less than a half of solar mass, the relevant density is less than $n_0$. The range of density probed by the future facilities like FAIR and NICA is quite relevant for the neutron star with a mass  of order of solar mass, where the central density is of order of a few times $n_0$~\cite{SPLE}.   The newly observed neutron star of $1.97$ sola mass is consistent with the higher central density with  the upper limit $\sim 10n_0$~\cite{1.97S}~\cite{LP}.   The symmetry energy and the effective masses of hadrons in medium (including koan) are the quantities depending on how strong interactions are implemented in the dense hadronic matter.

In this paper, we make a highly simplified discussion of how flavor symmetry can figure in the equation of state (EoS) that enters into the structure of compact stars, i.e., their maximum mass stable against gravitational collapse vs. their radii. Our central theme is that baryonic matter at densities exceeding that of normal nuclear matter at present is more or less unknown, and that one promising approach is to put skyrmions (in 4D) or instantons (in 5D) on crystal lattice~\cite{multifacet}. We anchor our reasoning on the hidden local symmetric nature of baryonic interactions that has been discussed since 1980's but recently rediscovered in gravity-gauge dual approach to the strong interactions where an infinite tower of hidden gauge fields figure.

We admit that given that the work which is technically and numerically involved is in an early stage, most -- if not all -- of what we find in this paper are far from rigorous, drawing ideas from qualitative or at best semi-quantitative results so far obtained in the WCU-Hanyang project. Even so, some are novel and promising.

The principal results we have obtained so far are as follows.

Our argument is largely based on the observation that when put on crystal lattice, a matter of multi-skyrmions (or multi-instantons in 5D) representing dense baryonic matter makes a topological phase change to a matter of half-skyrmions (or ``dyonic salts") at a density $n_{1/2}$ lying (not far) above the nuclear matter density $n_0$. While this statement is firm only in the $N_c$ limit, the topology change is likely to be robust and its physical implication could be independent of the crystal artifacts.

A prediction~\cite{SLPR,PLRS} that follows from the hidden local symmetric nature of the strong interactions that is distinct from other approaches is that the compressed baryonic matter is driven via ``symmetry mending"~\cite{mendedsymmetry} to an infrared fixed point called ``dilaton-limit fixed point" which lies at a density near but below chiral restoration. The consequence is that since the isovector channel involving $\rho$-NN coupling is affected, the standard RMF approach where the $\rho$-NN coupling controls the symmetry energy becomes a suspect.

One of the striking effects of the topology change is the modification of what was formerly known as ``BR (Brown-Rho) scaling " to a new scaling that will be referred, for reasons that will be made clear,   to as ``BLPR (Brown-Lee-Park-Rho) scaling."  This has a drastic consequence on the nuclear symmetry energy. Due to this topology change, the structure of nuclear tensor forces gets significantly  altered, which then determines the structure of the nuclear symmetry energy which is dominated by the tensor forces. The topology change also influences how hadrons behave on top of the dense background, such as the onset of kaon condensation and the appearance of hyperons at high density, a controversial issue for the EoS of compact stars. We suggest that kaon condensation and hyperon presence in compact-star matter can be treated on the same footing in a unified way. Our approach predicts that hyperons appear only after kaons condense in disagreement with the usual phenomenological mean field approaches where hyperons appear first as strangeness degree of freedom and then suppress or delay kaon condensation.

We propose that kaon condensation can be taken as a doorway to strange quark matter, a notion which seems to be consistent with the structure of the 1.97 solar mass star.\cite{KLR}

\section{Symmetry Energy}
\indent\indent The nuclear symmetry energy denoted $\epsilon_{sym}$ figures importantly in nuclear physics and in compact-star physics. Although it is more or less controlled by experiments up to near nuclear matter density $n_0$ -- and is probed slightly above $n_0$ in heavy-ion collisions, it is almost completely unknown at high density relevant to the interior of compact stars going up to, e.g., $\sim 10n_0$. There are a large number of theoretical predictions for $\epsilon_{sym}$ that range widely above $n_0$ but given the total absence of model-independent tools to gauge the reliability and experimental information, none of them can be taken seriously. In this section, we discuss the various aspects of the symmetry energy, some rigorous and some not, that have not been addressed in the literature.
\subsection{Isospin symmetry}
\indent\indent To start with, we review the elementary -- albeit text-book -- notion of the symmetry energy so as to define the language that we shall use for our analysis. Although there is nothing new in this discussion, it will sharpen our arguments to be developed in a way to enable us to address the issues raised in Introduction.

The (strong) interaction between  nucleons is charge symmetric and has been formulated in an $SU(2)$ symmetric way, i.e., isospin symmetry. Proton and neutron are assigned to be members of a doublet and all hadrons are classified into $SU(2)$ multiplets.

 Nuclei are the systems where many nucleons are bound together. Involving many-body correlations, it is difficult to describe them directly from the basic nucleon-nucleon interaction, not to mention from QCD.   Even when we are provided with an effective Lagrangian or Hamiltonian for nucleon-nucleon interactions,  it is not a straightforward task to calculate physical observables of many-body systems from it.

 Throughout this paper, we will ignore Coulomb interactions. Then the Lagrangian is $SU(2)$-symmetric, which is blind to flavor in the sense that it cannot  distinguish between proton and neutron.  The interaction Hamiltonian commutes with $SU(2)$ generators, $\vec{I}$,
 \be
 [I_i , H_{int}] = 0,
 \ee
 \be
 [I_i , I_j ] = i \epsilon_{ijk} I_k.
 \ee
 Thus the proton $(I_3=1/2)$ numbers and neutron $(I_3=-1/2)$ numbers are conserved, so that we can classify the eigenstate by the definite number of protons and neutrons $(I_3=\frac 12 (N_p-N_n)$):
 \be
 |N_p, N_n \ra\, .  \label{eigen2}
 \ee
 The energy of the eigenstate of the Hamiltonian does not depend on $I_3$ but on $I^2$.    The eigenstate, Eq.(\ref{eigen2}), can be decomposed into the irreducible representations (multiplets) of $SU(2)$, Clebsch-Gordan series,  as
 \be
 |N_p, N_n \ra =  \Sigma_I C_I |I: N_p, N_n\ra\label{CGseries}
\ee
where $ |\vec{I}|^2 = I(I+1)$. The energy of the state is given by
\be
 E(N_p, N_n)= \la N_p, N_n | H |N_p, N_n \ra  = \Sigma_I |C_I|^2  E_{I}
\ee
where $E_I$ is a reduced matrix element of the Hamiltonian, $H = H_0 + H_{int}$,
\be
E_I = \la I|| H ||I \ra
\ee
which is independent of $I_3$, and depends on the details of the strong interactions for each $I$-channel.

Although it may appear that the energy is independent of the compositions of protons and neutrons for a given total number of nucleons
\be
N= N_p + N_n,
\ee
different compositions have different decompositions into multipletes, i.e., different sets of $C_I$.
Therefore  different compositions of protons and neutrons lead to different energies.

Here are a few examples:
\begin{enumerate}
\item  The states with $ N_p$  and $N_n$ exchanged have the same set of $C_I$, so the energy is degenerate,
\be
E(N_p, N_n) = E(N_n, N_p).
\ee
\item The state with $N_p =N$ and $N_n=0$ has the highest $I_3$  but the state with $N_p =N/2$ and $N_n=N/2$ cannot have that state.  Moreover, the state with $N_p =N/2$ and $N_n=N/2$ can have an isosinglet component while the state with $N_p =N$ and $N_n=0$ cannot.     Therefore they should have different decompositions of multiplets or equivalently  different sets of   $C_I$.  Hence, in general,
\be
E(N_p=N, N_n=0) \neq E(N_p=1/2, N_n=1/2).
\ee
This is of course a well-known relation that explains why the nuclear symmetry energy appears in asymmetric nuclear matter. It is not a result of isospin symmetry breaking of the strong interactions. The strong interaction is isospin-symmetric, but we are considering the states with different Clebsch-Gordan decompositions.
\end{enumerate}

In short, the density dependence of nuclear symmetry energy can be understood in terms of the density dependence of $C_I(n)$ and $E_I(n)$.  The normalization,  $\Sigma_I |C_I|^2 =1$, implies that at higher density, the probability of having a specific $I$ becomes very small.  {However, since the total energy of the system depends on the product of $C_I(n)$ and $\la I||H_{int}|| I\ra$, the importance of a particular isospin-$I$ state depends on the nature of the interaction Hamiltonian.

Of course, the Pauli exclusion principle can be considered  as an isospin symmetric interaction, which forces  the quantum states to be totaly anti symmetric in permutation of identical particles.
In nuclear matter, basic constituents are protons and neutrons, which are subjected to Pauli exclusion principle.  To appreciate it transparently, let us assume that the interactions can be turned off, namely, free gas approximation. Since there is no interaction, the reduced matrix elements for $I$-channel lose the $I$-dependence which means that  there is no composition dependence of the energy:
\be
\hat{E}^{free}(N_p, N_n) =  \hat{E}^{free}(N)
\ee
where $\hat{E}$ is the energy of the free-gas system without Fermi statistics.
However, when we apply the Pauli principle to the system, the energy for the symmetric state ($N_p=N_n$) should be lower than the state with protons or neutrons only. So the composition dependence is present even in the free Fermi gas system:
\be
E^{free}(N_p=N, N_n=0) > E^{free}(N_p=1/2, N_n=1/2).
\ee
We will return to this below. This aspect together with the isospin-dependent nuclear force will be exploited for a transparent description of the symmetry energy.
\subsection{Nucleon-nucleon interactions}
\indent\indent For a system of $N$ nucleons, the energy differences of the states with different composition of protons and neutrons are encoded in the symmetry energy defined by subtracting the energy of the state with symmetric compositions, $N_p = N_n= N/2$, from the energy of system composed of $N_p$ protons and  $N_n$ neutrons,
\be
E_{sym}(N_p, N_n) \equiv  E(N_p, N_n) -  E(N_p=N/2 , N_n=N/2)
\ee
or
\be
E_{sym}(N,x)\equiv E(N,x) - E(N,x=1/2)
\ee
with $x = N_p/N_n$. Clearly $E_{sym}(N,x=1/2) =0$ and
\be
E_{sym}(N_p, N_n) = E_{sym}(N_n, N_p),
\ee
hence
\be
E_{sym}(N,x)= E_{sym}(N,1-x).
\ee
Expressed in terms of the energy density $\epsilon=E/V$, we have
\be
\epsilon(n,x) = \epsilon(n,x=1/2) + \epsilon_{sym}(n,x), \label{epsn}
\ee
where $n=N/V$ is the nucleon number density.
Written in the quadratic approximation used in the literature\footnote{Higher order corrections -- in $(1-2x)$ -- are known to be small near $n_0$ but could be substantial at high density, $\sim 7\%$ of the quadratic term $\epsilon_{sym}$ at $\rho\gsim 6n_0$\cite{cai-chen}.}
\be
\epsilon_{sym}(n,x) \approx  n S(n)(1-2x)^2. \label{sn}
\ee
{One can see this trivially holds for non-interacting systems, say, within 5 $\%$ accuracy, as given by
\be
\epsilon^{free}_{sym}(n,x) =  n S^{free}(n)(1-2x)^2, \label{sfn}
\ee
where
\be
S^{free} =  (2^{2/3} -1)\frac{3}{5}E_F^0(\frac{n}{n_0})^{2/3}, \label{sfreeapp}
\ee
and $E_F^0 = \frac{1}{2m}\left(\frac{3\pi^2 n_0}{2}\right)^{2/3}$.}

In what follows, we will focus on the ``symmetry energy factor" (or simply symmetry energy) $S(n)$
\be
S(n) = [\epsilon(n,0) - \epsilon(n,x=1/2)]/n.
\ee
Here we assume that Eq.~(\ref{sn}) holds\footnote{ There may be deviation from this in the presence of kaon condensation. See below.} for any value of $x$, for example see \cite{CPR}\cite{dkm}.

The overall structure of the symmetry energy can be inferred by considering the two extreme cases with  $x=0$ and $x=1/2$. The configuration with $x=0$ with no protons is governed by neutron-neutron interactions only, and similarly for $x=1$ with no neutrons but only with protons.
Consider two identical systems of neutrons only with nucleon number $N/2$ (or density $n/2$) for each. Define the sum of the two $\tilde{\epsilon}(n,x=1/2)$ given by
\be
\tilde{\epsilon}(n,x=1/2) \equiv 2 \epsilon(n/2,x=0).\label{tepsilon}
\ee
Now consider a nuclear matter consisting of $N/2$ neutrons and $N/2$ protons, for which  the energy density  is given by $\epsilon(n,x=1/2)$. Then $\Delta_{np}$ defined  by
\be
\Delta_{np}(n) &\equiv& \epsilon(n,x=1/2) -  \tilde{\epsilon}(n,x=1/2)\label{Deltanp} \\
               &=& \epsilon(n,x=1/2) - 2 \epsilon(n/2,x=0)
\ee
can be used as a measure of the $n-p$ interactions.
Then  we have for the symmetry energy for the $x=0$ configuration with density $n$
\be
\epsilon_{sym}(n,0) &=& \epsilon(n,0) - \epsilon(n,x=1/2), \\
&=& \epsilon(n,0) - 2 \epsilon(n/2,x=0) -  \Delta_{np}(n). \label{epsns}
\ee
Defining $\Delta_{nn}(n)$ as a measure of what could be called ``double-neutron" interactions by
\be
 \Delta_{nn}(n) = \epsilon(n,0) - 2 \epsilon(n/2,x=0), \label{Dltann}
\ee
we have
\be
\epsilon_{sym}(n,0) = \Delta_{nn}(n) - \Delta_{np}(n)\, .\label{epsymm}
\ee
From eq.(\ref{sn}), we have for the symmetry-energy factor
\be
S(n) = [\Delta_{nn}(n) - \Delta_{np}(n)]/n. \label{snd}
\ee
This is the expression we will use for our discussions that follow.

One can make a few simple observations with this formula. For the system of non-interacting nucleons, $\Delta_{np}(n)=0$, but $\Delta_{nn}(n) >0$ using  the free Fermi-gas approximation.  As is well known, this gives a positive symmetry energy factor $S$. Suppose that $n$-$p$ interactions are  attractive, i.e., $\Delta_{np}(n) <0$. Then {\it the ``attraction" in the interaction increases the symmetry-energy factor, that is, makes it effectively ``repulsive."} For those outside of the field, this aspect can cause confusion.\footnote{For instance, at normal nuclear density $n_0$, $\Delta_{nn}(n)$ includes the effects of ``repulsion" due to Pauli principle.  It gives $\sim 13$ MeV. The attractive n-p interaction, contributing with the opposite sign, gives the ``repulsion" $\Delta_{np}(n) \sim 18$ MeV, making the total $\sim 30$ MeV as is observed in nature. Roughly, $\Delta_{nn}(n)$ is determined by the Fermi-Dirac statistics and $\Delta_{np}(n)$ by the n-p interactions.  More quantitatively, however,  $\Delta_{np}(n)$ and $\Delta_{nn}(n)$ depend strongly on the details of the nucleon-nucleon interactions as we will illustrate next.}
\section{Anatomy of the Symmetry Energy}
\indent\indent There are many approaches to the symmetry energy in the literature which can be broadly categorized by (a) phenomenological and (b) effective field theoretic (EFT). We shall now use Eq.~(\ref{snd}) to unravel how nuclear forces control the symmetry energy.
\subsection{Phenomenological approaches}
\indent\indent As the first example,  we consider a phenomenological model for the energy density and symmetry energy factor $S$.\footnote{The variety of sophisticated nuclear many-body approaches -- including three-body forces -- with parameters  adjusted to available experimental data, can describe the EoS up to the density constrained by experiments. However going beyond $n_0$ involves extrapolations constrained neither by experiments nor by theory. The purely phenomenological model with adjustable parameters that is considered here encompasses such many-body approaches, so we will not discuss them here.} Eschewing going into details of pros and cons of various models, we take one simple approach called ``MID" (short for ``momentum-independent interaction") used by Li {\it et al.}~\cite{lichenko} referred below to as LCK. It has the generic form that illustrates our basic point. Even though by definition, phenomenological models mix in basic interactions with many-body correlations, so it is difficult to single out symmetry effects from dynamics, it is feasible to see how they enter into the symmetry energy. For this, we rewrite Eqs.~(5.1) and (5.5) of \cite{lichenko} as
\be
\epsilon^{int}(n,x=1/2) &=& n[\frac{A}{2}  \frac{n}{n_0} + B  (\frac{n}{n_0})^{\gamma}], \label{be}\\
S^{int}(n) &=& F_{\alpha} \frac{n}{n_0} + (18.6-F_{\alpha})(\frac{n}{n_0})^{C_{\alpha}}, \label{bsn}
\ee
where $A, B, C_\alpha, F_\alpha,\gamma$ are the parameters to be determined by experimental data. The energy density $\epsilon$ and symmetry energy factor (or symmetry energy in short) $S$ are now given by
\be
\epsilon(n,x) &=& n [m_N + \frac{3}{5}  E_F^0(\frac{n}{n_0})^{2/3} + \frac{A}{2}  \frac{n}{n_0} + B  (\frac{n}{n_0})^{\gamma} \nonumber \\
      && +(1-2x)^2 S(n)], \label{1} \\
S(n) &=& (2^{2/3} -1)\frac{3}{5}E_F^0(\frac{n}{n_0})^{2/3} + F_{\alpha} \frac{n}{n_0} + (18.6-F_{\alpha})(\frac{n}{n_0})^{C_{\alpha}}. \label{2}
\ee
The parameters $A$, $B$ and $\gamma$ determined by experiments are $A= -298.3$ MeV, $B= 110.9$ MeV, $\gamma = 1.21$ and $E_F^0 = 37.2$MeV. The parameters $F$ and $C$, however, cannot be uniquely determined. Thus there are possible values for $F$ and $C$ labeled by $\alpha$ (denoted $x$ in \cite{lichenko}).

Then using (\ref{Deltanp}) and (\ref{Dltann}), we get
 \be
 \Delta_{nn} &=& n[(2^{2/3} -1)\frac{3}{5}E_F^0(\frac{n}{n_0})^{2/3} \nonumber \\
 & & + \frac{1}{4}A(\frac{n}{n_0}) + B  (\frac{n}{n_0})^{\gamma}(1-2^{-\gamma}) +(1-1/2) F_{\alpha}(\frac{n}{n_0}) \nonumber \\
 & & + (18.6-F_{\alpha})(\frac{n}{n_0})^{C_{\alpha}}(1-2^{-C_{\alpha}})] \label{Dnnlck}, \\
 \Delta_{np} &=& n[\frac{1}{4}A(\frac{n}{n_0}) + B  (\frac{n}{n_0})^{\gamma}(1-2^{-\gamma}) -\frac{1}{2} F_{\alpha}(\frac{n}{n_0}) \nonumber \\
 & & -  (18.6-F_{\alpha})(\frac{n}{n_0})^{C_{\alpha}}2^{-C_{\alpha}}]. \label{Dnplck} \ee
We can see the nature of degenerate Fermi gas only appears in $\Delta_{nn}$ as expected from  the forms of the parametrization for the energy of the symmetric matter and the symmetry energy.

The results for $n=n_0$ for three values of $\alpha$'s, $\alpha= -1, 0, 1$, are summarized in Table~\ref{Delta}.
\begin{table}[h]
\begin{center}
\begin{minipage}{.8\textwidth}
\begin{center}
\begin{tabular}{l | c | c | c}
\hline
\hline
$\alpha$ & $-1$ & $0$ & $1$\\
\hline
\hline
$\Delta_{np}/n|_{n_0}$ & $-18.5$ & $-23.1$ & $-27.9$ \\
\hline
$\Delta_{nn}/n|_{n_0}$  & $13.2.$ & $8.6$   & $3.8$\\
\hline
\hline
\end{tabular}
\caption{$\Delta$ in MeV}\label{Delta}
\end{center}
\end{minipage}
\end{center}
\end{table}

Recalling that $\Delta^{free}_{nn}/n|_{n_0} = 13.1$ MeV for the free nucleon system without interactions, we see that while the attraction in the n-n attraction is negligible for $\alpha =-1$, it becomes much stronger for $\alpha=1$, with the exclusion repulsion largely compensated by the attraction.  As for  $\Delta_{np}$, the n-p attraction becomes stronger, the greater the $\alpha$ parameter.
\begin{figure}[ht]
\begin{center}
\includegraphics[width=4.7cm]{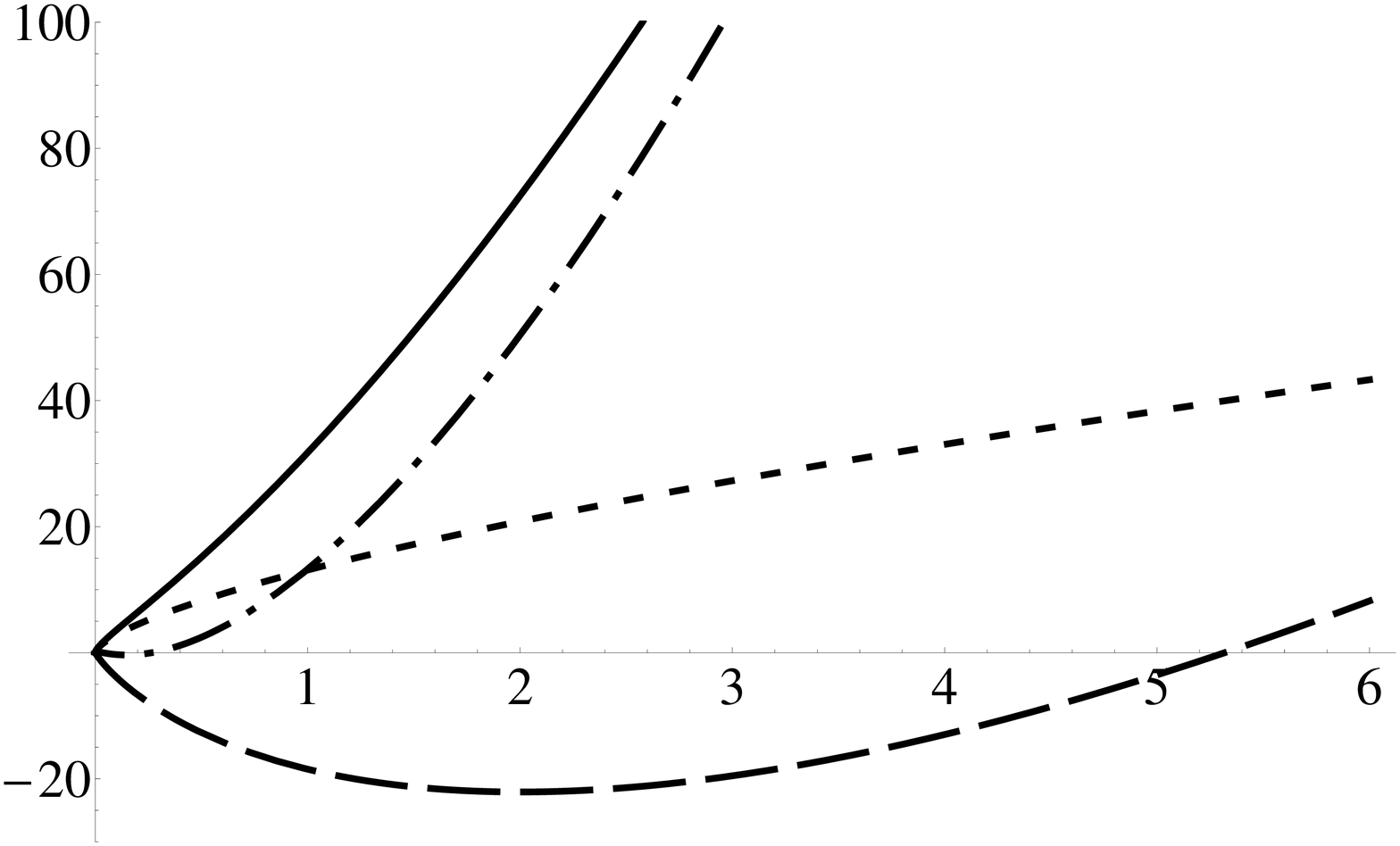}\includegraphics[width=4.7cm]{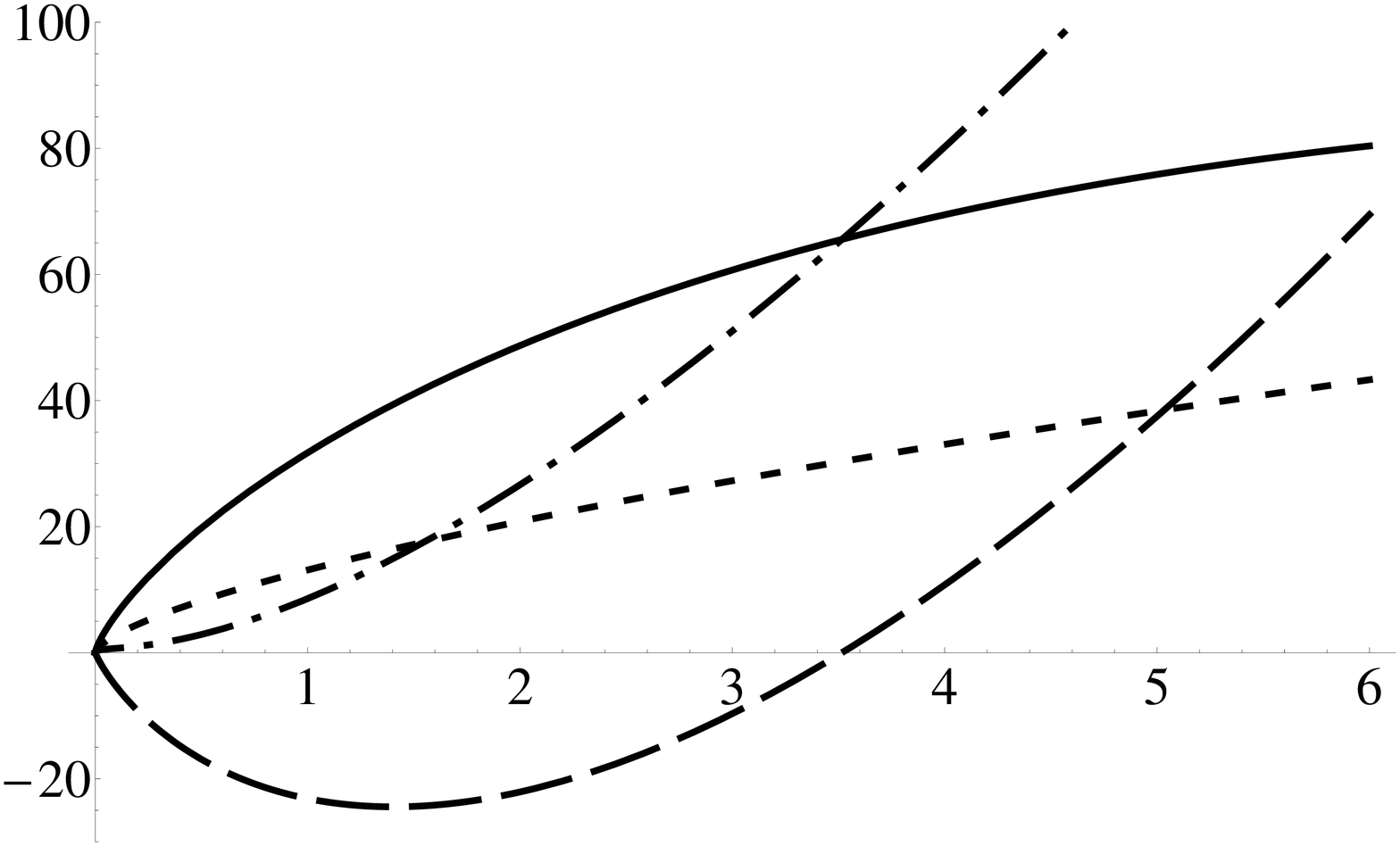}
\includegraphics[width=4.7cm]{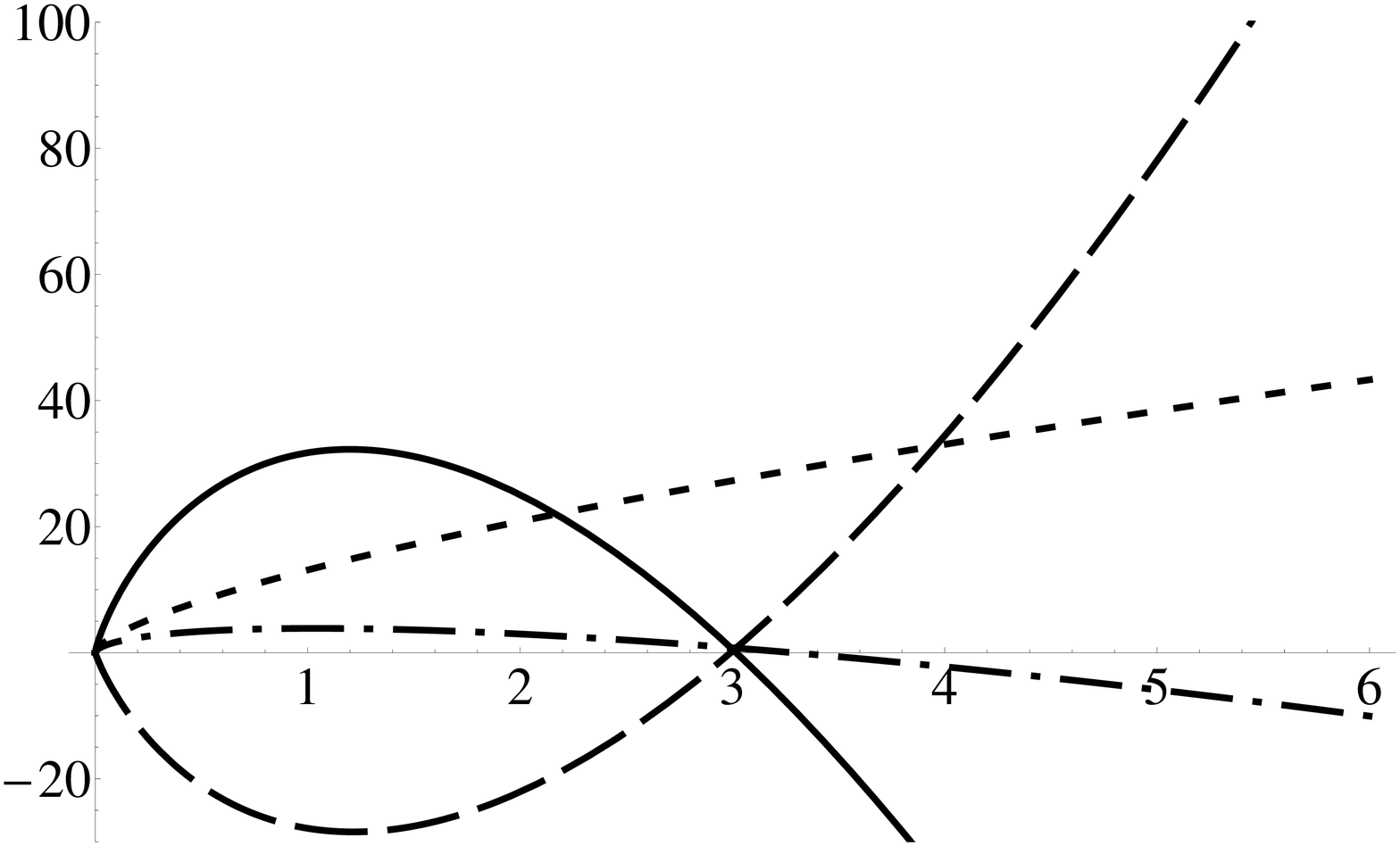}
\caption{The symmetry energy factor $S(n)$ (thick line), $\Delta_{np}/n$ (dashed), $\Delta_{nn}/n$(dash-dotted) and the symmetry energy for non-interacting nucleons (dotted) for the LCK model with $\alpha=-1$ (left panel), $=0$ (center panel) and $=+1$ (right panel) referred to as ``supersoft." The vertical axis is in unit of MeV and the horizontal axis is $n/n_0$.}
\label{a-1}
\end{center}
\end{figure}

The density dependence of $\Delta_{nn}$ and $\Delta_{np}$ are shown in Fig.~\ref{a-1} for $\alpha=-1,0,1$. These results illustrate how strong interactions via nuclear forces can influence the $S$ -- an indispensable part of the EoS, giving an answer posed in Introduction. While fine-tuned to fit experiments up to $\sim n_0$, the structure of $\alpha=1$ for n-n interaction is basically different beyond $n_0$ from both $\alpha=0$ and $\alpha=-1$. As we will discuss below, the key role here is played by the nuclear tensor force. What governs $S$ above $n_0$ has therefore a huge influence on the structure and the fate of compact stars.
\subsection{Relativistic mean-field approaches}\label{RMF}
\indent\indent An alternative to the strict phenomenological approach anchored on standard nuclear many-body approach is to resort to phenomenological Lagrangians that incorporate {\em as complete} relevant degrees of freedom as feasible for practical calculations. One writes an effective Lagrangian in the spirit of doing effective field theory and treat it in {\em the mean field} ignoring higher-order loop corrections embodied in chiral perturbation theory, and fixing the parameters of the Lagrangian from experiments done in zero-density vacuum. We are therefore ignoring possible nonlocality associated with loop graphs. The Lagrangian is chosen on the basis of ``naturalness" imposed on high order terms consistent with the power counting prescribed by the symmetry involved, chiefly chiral symmetry. The basic premise of such an approach is that in doing nuclear physics, the mean-field theory with an effective Lagrangian with correct symmetries is equivalent to doing Landau Fermi-liquid theory for nuclear matter~\cite{matsui} and that such a mean field theory can be extrapolated to high density regime relevant to compact stars. We shall call this ``standard relativistic mean field approach" (SRMFA for short) to distinguish it from the mean-field of the hidden local symmetry (HLS for short) Lagrangian that we will discuss later. In the next subsection we shall propose that hidden local symmetric approaches implemented with both scalar and baryonic degrees of freedom and appropriate scaling of the parameters is more predictive when treated in mean field.

Here we will briefly illustrate what one can expect and what one has found in standard phenomenological mean-field approaches. In considering symmetric and asymmetric nuclear matter in mean field, we restrict to two flavors and ignore the pion field.  The relevant degrees of freedom for the case at hand are the nucleons $N$, the $\rho$ and $\omega$ mesons and a low-lying scalar that we will denote $s$. It is important that the $s$ is identified as a chiral scalar, not the fourth component of the chiral four vector usually denoted $\sigma$. The leading Lagrangian with the lowest-dimension field operators take the usual form which can be read off directly from the hidden local symmetric Lagrangian (\ref{bdHLS}) in Section \ref{hlslag} that contains both the dilaton scalar $\chi$ -- which we will identify later with $s$ -- and the nucleon field $N$. What is not found in (\ref{bdHLS}) but added in SRMF are higher-dimension field operators of the form
\be
\delta {\cal L} (N,S,\omega_\mu,\rho_\mu)&=&b_S (g_s s)^3 +c_\sigma (g_s s)^4 +c_\omega (g_\omega^2\omega_\mu\omega^\mu)^2\nonumber\\
&& +g_\rho^2\rho^a_\mu\rho^{a\mu} (d_S g_s^2 s^2 +d_V g_\omega^2 \omega_\mu\omega^u)+\cdots \label{highdimension}
\ee
where $g_{s,\omega,\rho}$ and $b,c,d$ are arbitrary constants fit to experiments. As an illustration, we take the parametric forms discussed by Piekarewicz and Centelles~\cite{PC} up to second order in the density fluctuation. The energy density of the asymmetric matter is written as
 \be
 \epsilon_{srmf}(n,x) = n[m_N + \epsilon_0 + \frac{1}{2} K_0 \tilde{u}^2 +S (1-2x)^2]\label{rmf}
\ee
where the symmetry energy factor is
\be
 S = J + L\tilde{u} + \frac{1}{2} K_{sym} \tilde{u}^2\, . \label{srmf}
 \ee
Here the quantities $J$, $L$ and $K_{sym}$ are given in terms of the parameters of the Lagrangian and mean-field values of $S$, $\rho_0$ and $\omega_0$ -- the explicit forms of which we need not write down, $\tilde{u} \equiv \frac{(n-n_0)}{3n_0}$, and $n_0 = 0.148$ fm$^{-3}$\footnote{This is somewhat smaller than the canonical value 0.16 fm$^{-3}$.} corresponds to the equilibrium density. In this parametrization, the kinetic parts are not explicitly shown in contrast to LCK. The parameters, $J,L,K_{sym}$, the energy density of the symmetric matter $\epsilon_0$ and the compression modulus $K_0$ are given in Table~\ref{rmfp} for two models ``FSU" and ``NL3" both of which are fit to reproduce what is given by experiments up to $\sim n_0$.
\begin{table}[h]
\begin{center}
\begin{minipage}{.8\textwidth}
\begin{center}
\begin{tabular}{l | c | c | c |c | c }
\hline
\hline
Model & $J$   & $L$  &  $K_{sym}$ & $K_0$ & $\epsilon_0$\\
\hline
FSU   & $32.59$ & $60.5$ & $-51.3$  & $230$ & -16.30\\
NL3   & $37.29$ & $118.2$ & $100.9$ & $271.5$ &-16.24\\
\hline
\hline
\end{tabular}
\caption{Bulk parameters characterizing the energy of symmetric nuclear matter
and symmetry energy~\cite{PC}. $K_0$ and $\epsilon_0$ are, respectively, the compression modulus and the energy density of the symmetric matter at its equilibrium.}\label{rmfp}
\end{center}
\end{minipage}
\end{center}
\end{table}
From (\ref{rmf}) and (\ref{srmf}), we find
 \be
 \Delta_{nn} &=& n[ \frac{1}{2} K_0 (\tilde{u}^2 - \bar{u}^2)  +  L(\tilde{u}-\bar{u}) + \frac{1}{2} K_{sym} (\tilde{u}^2- \bar{u}^2)] \label{delrmfnn} \\
 \Delta_{np} &=& n[  \frac{1}{2} K_0 (\tilde{u}^2- \bar{u}^2)
  - (J + L\bar{u} + \frac{1}{2} K_{sym} \bar{u}^2)],
 \label{delrmfnp}
 \ee
where $ \bar{u} \equiv \frac{(n/2-n_0)}{3n_0}$.   The resulting symmetry energy is shown in Fig.~\ref{meanfield} for the parametric forms of FSU and NL3.
\begin{figure}[ht]
\begin{center}
\includegraphics[width=5.0cm]{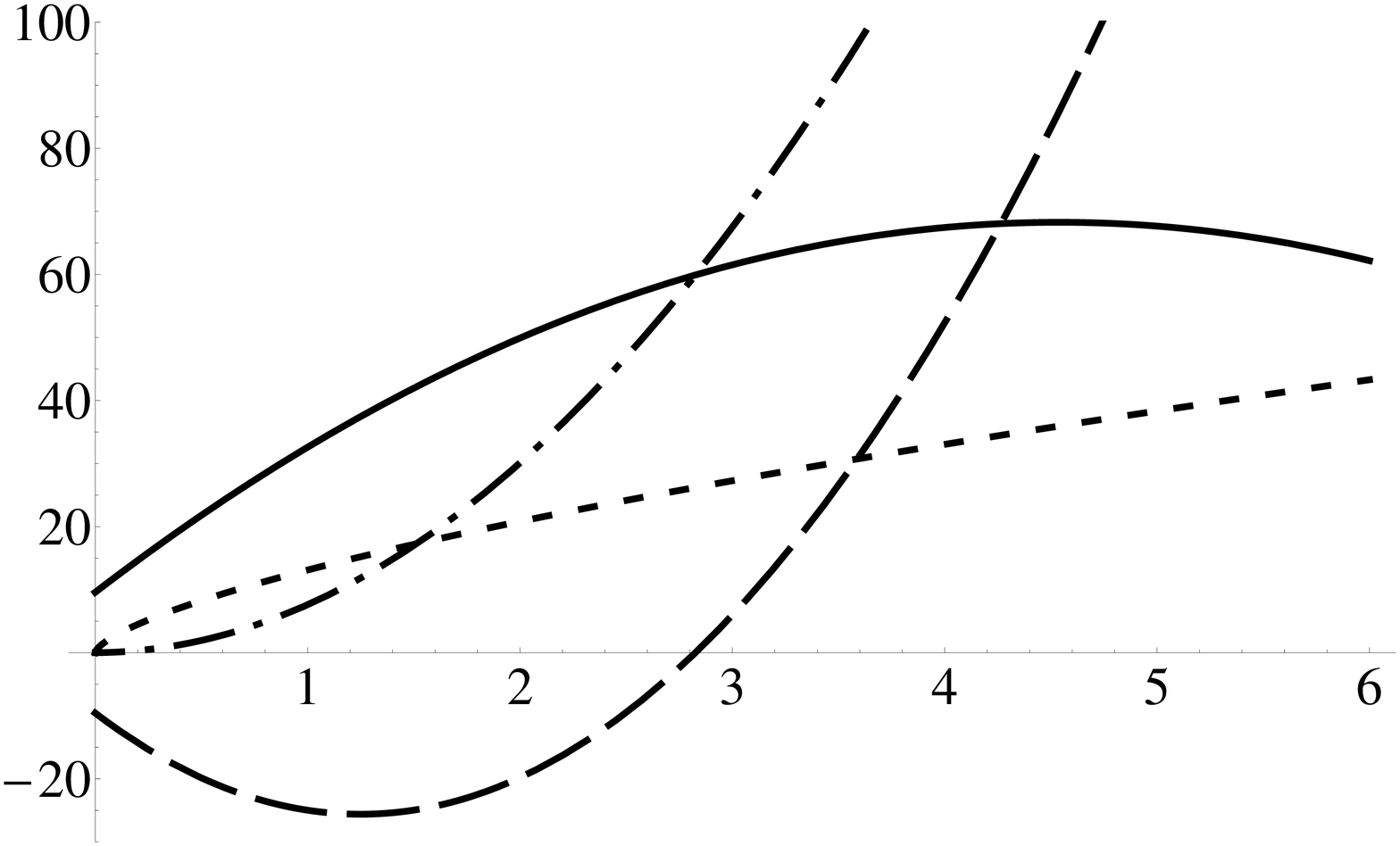}\includegraphics[width=5.0cm]{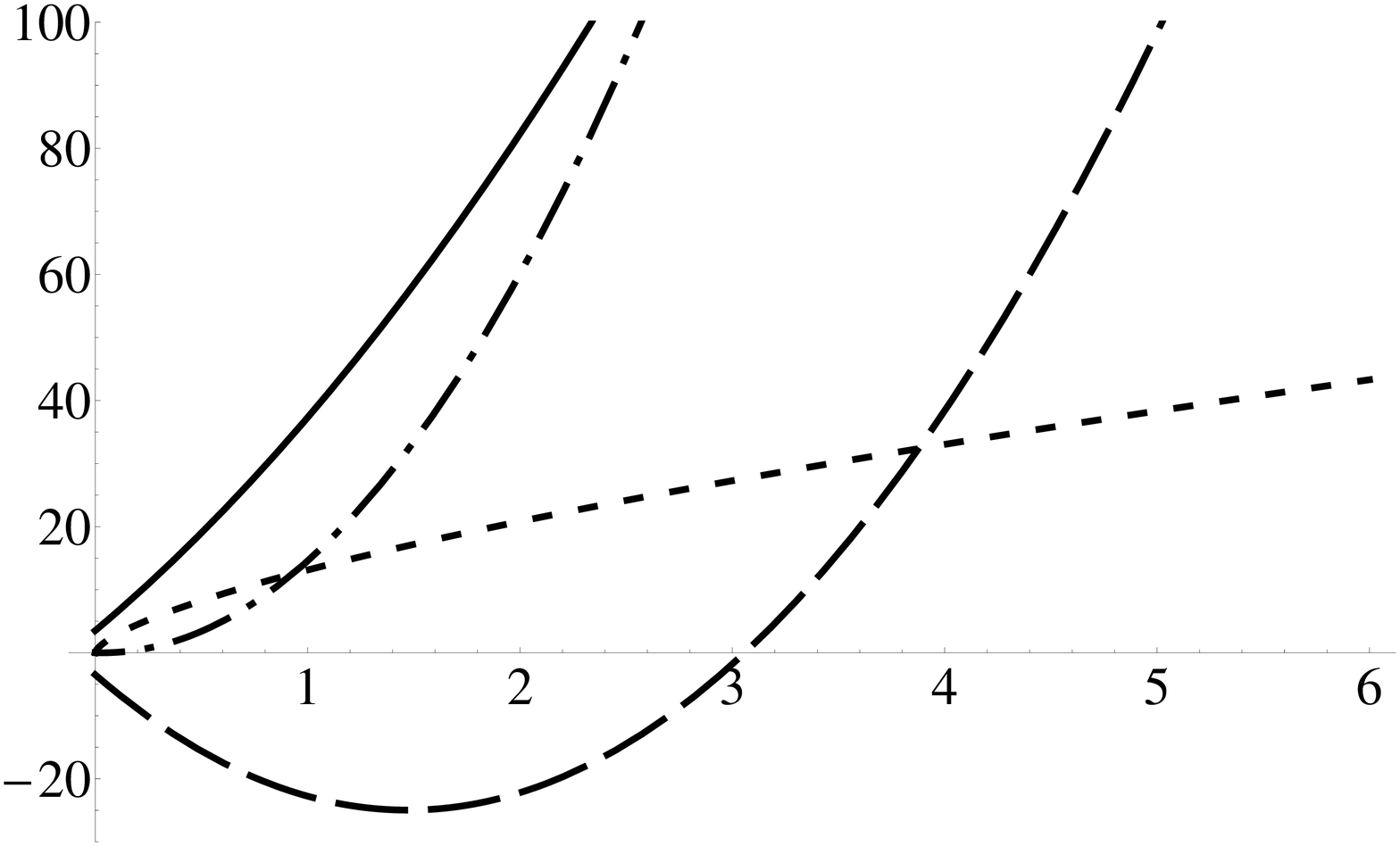}
\caption{Same as Fig.~\ref{a-1} for the FSU model (left panel) and the NL3 model (right panel).}
\label{meanfield}
\end{center}
\end{figure}

Put together with the results of the standard phenomenological nuclear models, one can say that while constrained by nature up to $\sim n_0$, the symmetry energy for $n>n_0$ is totally undetermined. {There is however one common feature in all these approach and it is that the attractive n-p interactions ($\Delta_{np} <0$) become repulsive ($\Delta_{np} >0$) at higher densities as seen in Figs.~\ref{a-1} and \ref{meanfield}.}

\subsection{Effective field theory approaches}
\subsubsection{\it Hidden local symmetry}\label{hlslag}
\indent\indent Given the wildly diverging predictions by the available models, it is clearly desirable to devise theoretical controls on the relevant quantities. At present no model-independent field theoretic approach is available for dense baryonic matter\footnote{There are attempts to apply chiral perturbation theory (B$\pi$ChPT) with baryonic chiral Lagrangian (involving nucleons and pions only) to nuclei and nuclear matter.  B$\pi$ChpT has some modest success in nuclei and nuclear matter but does not work for some quantities that are intricately tied to Fermi-liquid fixed-point structure of nuclear matter such as the orbital gyromagnetic ratio $\delta g_l$ for which mean-field approach works better\cite{friman-rho}. In particular, it is not likely to work for dense matter we are concerned with.}: lattice QCD cannot access that regime. The approach closest in spirit to -- though not directly derived from -- QCD is the skyrmion description of dense baryonic matter, and the first attempt to deduce the symmetry energy from it was made in \cite{LPR-halfskyrmion}. Anchored on large $N_c$ properties of QCD, the approach captures nonperturbative aspects of large $N_c$ gauge structure of the strong interactions -- and most importantly for any number of $A$ baryons for $A=1, ..., \infty$ with $A$ identified with the topological winding number $B$ of the soliton. While the skyrmon picture was introduced by Skyrme for baryons~\cite{skyrme}, the concept has proven to be extremely powerful and pervasive in many areas of physics, most successfully in condensed matter physics~\cite{multifacet}.

In applying this concept to dense baryonic matter, however, there is a technical difficulty and that is, it is not known how to write down an effective Lagrangian that captures QCD in a manageable form. Written in the most minimal way in the large $N_c$ limit is the Skyrme Lagrangian that contains pions only, consisting of the two-derivative current algebra term and the quartic term (called Skyrme term) that provides topological stability to the skyrmion. There is nothing in principle that prevents from bringing in higher derivative terms as the energy scale is increased but one cannot work with such a Lagrangian that involves too many unknown parameters to control.

An alternative field theoretic approach is density-functional approach anchored on mean-field treatments of effective Lagrangians with pertinent symmetries taken into consideration. This approach is essentially related to the standard relativistic mean-field (SRMF) theory mentioned in Section \ref{RMF}. Here again the plethora of higher-order terms involved in the approach are difficult to control systematically.

One way to circumvent this difficulty in both approaches is to bring in vector degrees of freedom that figure at higher mass scales. This can be done most effectively by exploiting hidden local symmetry which can appear in the leading $N_c$ order on the same footing with the pion. This can be done naturally by exploiting the redundancies inherent in the chiral field\footnote{In this paper we will confine to two flavor $SU(2)$.} $U=e^{2i\pi/F_\pi}\in SU(N_f)_L\times SU(N_f)_R/SU(N_f)_{V=L+R}$, namely $U=\xi_L^\dagger\xi_R=\xi_L^\dagger g^\dagger (x)g(x)\xi_R$ where one has $g(x)\in SU(N_f)_V$. Gauging the redundancy $g(x)$ by introducing the gauge fields $V$ makes the nonlinear sigma model Lagrangian extended to the energy scale of $V$. The gauge fields here can be considered as ``emergent" from pion interactions going up from the current algebra scale. For $N_f=2$, the lowest-lying hidden (or ``emergent") gauge mesons are $\rho$ and $\omega$, so we will be concerned with $V=\rho, \omega$.

The power of hidden local symmetry is that it allows a systematic chiral counting, hence a systematic chiral perturbation expansion with vector mesons included. Furthermore it can consistently treat the situation where the vector meson masses become as light as the pion mass -- which is massless in the chiral limit, a feature which is lacking in phenomenological models with no local gauge symmetry~\cite{HY:PR}.

Another power in this approach is that an infinite tower of vector mesons can be incorporated as hidden gauge fields. This can be derived bottom-up by exploiting the infinite redundancies and generating the infinite tower of hidden gauge fields, yielding a de-constructed 5D Yang-Mills action~\cite{son-stephanov}. It can also be obtained top-down by reduction via branes from string theory. Dual -- and closest -- to QCD with correct chiral symmetry is the Sakai-Sugimoto hQCD model~\cite{SS}.

What the impact of the higher tower of vector mesons on RMF approach is has not yet been investigated. On the other hand, for the skyrmion approach, it is found that the tower of HLS fields encoded in the 5D YM action markedly improve on the structure of skyrmion for single baryon as well as for many-baryon systems~\cite{sutcliffe}. By taking a flat space 5D YM action and using the BPS skyrmion, Sutcliffe shows that the more gauge fields incorporated, the better the skyrmion description of the baryonic systems. This implies that broken scale symmetry tends to be restored as more vector mesons are incorporated into the skyrmion structure. We will see that this feature emerges in the skyrmion crystal we consider in this paper.
\subsubsection{\it HLS with baryons and dilaton}
\indent\indent In describing nuclei and nuclear matter, a low-lying scalar meson degree of freedom, either as a local field or composites, is essential. It gives the necessary intermediate-range attraction needed for the binding of the system. It is however missing in HLS with or without the infinite tower. How to incorporate scalar fields in hidden local symmetric Lagrangian is problematic because of the unnaturalness associated with scalar fields and has not been worked out satisfactorily in that gauge theory. Focusing on mean-field approaches, we choose to introduce the needed scalar degree of freedom via the trace anomaly of QCD and introduce the scalar as the dilaton that accounts for the ``spontaneously broken" scale invariance {\it in the presence} of the explicit breaking due to the anomaly. How this can be done is explained in \cite{LR-dilaton}. The key assumption in \cite{LR-dilaton} is that there is a ``soft" dilaton $\chi_s$ that is locked to chiral symmetry the $vev$ of which vanishes when chiral symmetry is restored and a ``hard" dilaton $\chi_h$ whose $vev$ remains non-vanishing above the chiral critical point, playing the role of an explicit symmetry breaking to scale symmetry in the low-energy (confinement) sector as required to make the spontaneous symmetry breaking viable~\cite{freund-nambu}. We identify $\chi_s$ as the pseudo-Goldstone scalar in the spontaneous breaking of scale symmetry.

One can readily write down such a dilaton-implemented HLS (dHLS for short) Lagrangian~\cite{SLPR}. Retaining only the lowest vector mesons $\rho$ and $\omega$, and integrating out the ``hard" dilaton, the dHLS has the form\footnote{ Considering $N_f=2$, we write the Lagrangian in the  $U(2)$ symmetric way with the vector field defined as $gV_\mu=g(\frac 12 \vec{\tau}\cdot\vec{\rho}_\mu +\frac 12\omega_\mu)$. The $U(2)$ symmetry seems to work well in the zero-density vacuum, but there is no good reason for it to be a good symmetry in medium. We will however assume $U(2)$ symmetry up to the density at which skyrmions make the transition to half-skyrmions. For the symmetry $SU(2)\times U(1)$, $gV_\mu$ should be replaced by ($\frac{g_\rho}{2}\vec{\tau}\cdot\vec{\rho}_\mu + \frac{g_\omega}{2}\omega_\mu$). }
\be
{\mathcal L}_{dHLS}
&=& {\mathcal L}_M
{}+ {\mathcal L}_\chi\,,\label{dHLS}
\\
{\mathcal L}_M
&=& \kappa \chi^2\, \mbox{tr}
\left[ \hat{\alpha}_{\perp\mu}\hat{\alpha}_\perp^\mu \right]
{}+ a \kappa \chi^2\, \mbox{tr}
\left[ \hat{\alpha}_{\parallel\mu}\hat{\alpha}_\parallel^\mu \right]
{}- \frac{1}{2g^2}\mbox{tr}\left[ V_{\mu\nu} V^{\mu\nu} \right]+
\cdots\,,\nonumber
\\
{\mathcal L}_\chi
&=& \frac{1}{2}\partial_\mu\chi \cdot \partial^\mu\chi
{}+ \frac{\kappa m_\chi^2}{8 F_\pi^2}
\left[ \frac{1}{2}\chi^4
{}- \chi^4\ln\left( \frac{\kappa\chi^2}{F_\pi^2}\right)\right]\,,\nonumber
\ee
where we write $\chi$ for $\chi_s$, $a = (F_\sigma/F_\pi)^2$ and $\kappa=F_\pi^2/F_\chi^2$, and $m_\chi$ is the mass of the soft dilaton. The quantities $\hat{\alpha}_{\perp\mu}$ and $\hat{\alpha}_\parallel^\mu$ are defined in the review \cite{HY:PR} and in \cite{SLPR}. The ellipsis stands for higher order terms that result by integrating out the higher members of the tower.

We will consider baryons arising from this Lagrangian or rather a truncated version of it as solitons in what follows, that is, in putting skyrmions on crystals. But it is interesting to put baryons explicitly into the Lagrangian as one does in baryon chiral perturbation theory. At a given order, the two ways of treating baryons, as skyrmions or local baryon fields, is equivalent and this equivalence was used also for describing intanton bayons in the SS'  hQCD model~\cite{HRYY,SSbaryon}. With both the dilaton field and the baryon (nucleon in this case) field, the Lagrangian has the form~\cite{SLPR}
\be
{\mathcal L}_{bdHLS}
= {\mathcal L}_N + {\mathcal L}_M
{}+ {\mathcal L}_\chi\,,\label{bdHLS}
\ee
with
\be
{\mathcal L}_N
= \bar{N}i\Slash{D}N
{}- \frac{\sqrt{\kappa}}{F_\pi} m_N \bar{N}N\chi
{}+ g_A \bar{N}\Slash{\hat{\alpha}}_\perp\gamma_5 N
{}+ g_V \bar{N}\Slash{\hat{\alpha}}_\parallel N+\cdots\,.
\ee

${\mathcal L}_M$ and ${\mathcal L}_\chi$ are as defined above. This is for flavor $U(2)$. For $SU(2)\times U(1)$,  $g_V \bar{N}\Slash{\hat{\alpha}}_\parallel N$ should be replaced by $(g_{V\rho} \bar{N}\Slash{\hat{\alpha}}_\parallel N -g_{V\omega}\bar{N}\frac{\Slash\omega}{2} N)$.
\subsubsection{\it bdHLS and the dilaton-limit fixed point}
\indent\indent We will discuss mean-field properties of the Lagrangian (\ref{bdHLS}) below. Here we consider how it can describe {\em both} low-density matter -- near the density of nuclear matter -- {\em and} high density matter near chiral restoration. For  suitably adjusted parameters such as the dilaton and vector-meson masses fit to data, we can employ the Lagrangian (\ref{bdHLS}) in mean field to give nuclear matter. This is essentially Walecka linear mean-field model. With the parameters endowed with appropriate density dependence~\cite{BR91}, it can actually describe fairly well all nuclear matter properties~\cite{song}. The reason for this is that the mean-field theory so formulated is equivalent to Landau Fermi-liquid fixed point theory and therefore it should work well very near the nuclear saturation point. However there is no reason to expect that this theory extended in a naive way to high density would work. In fact near the chiral transition, the scalar dilaton which is a chiral scalar cannot be the relevant scalar. It must instead be the fourth component $\sigma$ of the chiral four vector $(\pi_1,\pi_2,\pi_3,\sigma)$. How the chiral scalar can be transmuted to $\sigma$ by density is unknown and needs to be clarified.

In \cite{SLPR}, the solution was found in terms of what is called ``dilaton limit" by Beane and van Kolck~\cite{beane-vankolck}. In what follows in this subsection, we will confine to the flavor $SU(2)$ sector of vector mesons, i.e., $\rho$ mesons, returning to the $U(1)$ sector later\footnote{Considering $U(2)$ symmetry in \cite{SLPR} is not justified in medium since there is no reason to expect that the symmetry holds in medium as it does in free space. The discussion in \cite{SLPR} on the suppression of repulsion due to $\omega$-exchanges therefore may be incorrect.}.  Combining the dilaton field $\chi$ and the chiral field $U$, redefine the fields as
\begin{eqnarray}
\Sigma
&=& U\chi\sqrt{\kappa} = \xi_L^\dagger\xi_R \chi\sqrt{\kappa}
= s + i\vec{\tau}\cdot\vec{\pi}\,,
\\
{\mathcal N}
&=& \frac{1}{2}\left[ \left( \xi_R^\dagger + \xi_L^\dagger \right)
{}+ \gamma_5\left( \xi_R^\dagger - \xi_L^\dagger \right) \right] N\,,
\end{eqnarray}
with the Pauli matrices $\vec{\tau}$ in the isospin space.
Expressed in these fields, the ``linearized" Lagrangian includes terms which generate singularities, negative powers of $\mbox{tr}\left[ \Sigma\Sigma^\dagger\right]$ that goes to zero in chiral symmetric phase. Those terms appear as multiplicative factors:
\begin{equation}
X_N = g_{V\rho} - g_A\,,
\quad
X_\chi = 1-\kappa\,.\label{dl}
\end{equation}
Here $g_{V\rho}$ represents the ``induced" $\rho$-N coupling replacing the $g_V$ relevant to $U(2)$ symmetry in \cite{SLPR}.
In order to prevent the divergences, one is then required to impose that as $\mbox{tr}\left[ \Sigma\Sigma^\dagger\right]\rightarrow 0$,
\be
X_N\rightarrow 0, \ \ X_\chi\rightarrow 0.
\ee
We assume that the large $N_c$ approximation is reliable for $g_A$ at large density~\footnote{This is seen in lattice calculations. It is found that $g_A$ calculated in quenched approximation in lattice QCD -- which corresponds to the large $N_c$ limit -- is close to what is found in unquenched calculations.}, which would imply that $g_A\rightarrow 1$ as the density increases toward chiral transition. At the point
\be
\kappa=g_A=g_{V\rho}=1\label{DL}
\ee
the bdHLS Lagrangian goes over to Gell-Mann-L\'evy sigma model. This is referred to as ``dilaton limit" (DL for short). The direct vector-meson coupling to the baryon in (\ref{bdHLS})
\be
g_{VN}\equiv g_\rho(1-g_{V\rho})
\ee
where $g_\rho$ is the $SU(2)$ hidden gauge coupling therefore goes to zero
\be
g_{VN}\rightarrow 0.\label{decoupling}
\ee
Thus approaching the DL, the vector mesons decouple at the tree level from baryons although $V$-$\pi$ couplings remain.

An interesting possibility with this limit is that it can be a fixed point of the bdHLS theory. Let us call it dilaton-limit fixed point (DLFP for short) and the density at which the fixed point is reached will be denoted $n_{dl}$. Indeed it has been shown that DL is an immediately visible fixed point in the one-loop RGE of (\ref{dHLS})~\cite{PLRS}, This theory may have a variety of other fixed points as in the original HLS theory~\cite{HY:PR} out of which the vector manifestation fixed is the only fixed point matched to QCD.  {\it Now we conjecture that approaching the Gell-Mann-L\'evy sigma model is tantamount to approaching the chiral restoration point, and that the DL is the fixed point of QCD in the same spirit as the VM fixed point is.} It remains to show that there are no other fixed points in this bdHLS theory (\ref{bdHLS}) that can be matched to QCD, in the same vein as that there are no other fixed points than the VM in HLS that match to QCD~\cite{HY:PR}.

One interesting question that remains unanswered is whether the DLFP and the VM fixed point in HLS theory are at the same density or even related to each other. It has been established that the VM fixed point remains intact in the presence of fermions whose dynamical mass goes to zero (in the chiral limit) at the chiral restoration point~\cite{HKimR}. This would be the case if baryons were chiral with masses generated dynamically. It seems highly likely that it will be the same if the baryon can be treated as heavy near the chiral transition point -- as in the mirror assignment for the baryons with a large chiral invariant mass~\cite{SLPR} -- in which case the fermion will decouple when approaching the VM point. It is not obvious, however, what would happen if the fermion were neither light nor heavy. Although much less studied, the situation seems quite similar for the DLFP (\ref{DL}) as shown in \cite{PLRS}. As we will see, our analysis suggests that they cannot be too far apart.

\subsubsection{\it Half-skyrmion matter and BLPR scaling}
\indent\indent We now turn to the skyrmion description of dense matter -- which is our principal tool for our problem. A work is presently in progress~\cite{hlmopry} but no results are yet available on skyrmions in dense matter anchored on the dilaton-HLS Lagrangian advocated above. We shall therefore pick a vastly simplified model and extract relevant information from it in applying to nuclear and dense matter. We will not be able to obtain numerically precise results, but we consider them to be qualitatively robust and then develop a procedure to make quantitative calculations.

The model we shall use is the Skyrme model that includes the soft dilaton field. We shall put the skyrmions of this model on the crystal and deduce the mean field features of the ground state of baryonic matter.

We first recall what one obtains in the absence of the dilaton, that is, the pure Skyrme model. From the skyrmions put on an FCC crystal with the chiral field described by  Atiyah-Manton's holonomy ansatz~\cite{atiyah-manton}, we learn~\cite{park-am} that in the half-skyrmion phase to which skyrmions make transition at a density $n_{1/2}$, scale invariance, broken in the skyrmion phase, gets restored. This can perhaps be understood as resulting from a ``symmetry enhancement" associated with lattice translation~\cite{manton-sutcliffe-book}. Here scale symmetry and chiral symmetry are connected in such a way that the restoration of scale symmetry is locked to the vanishing of the chiral condensate $\la \bar{q}q\ra$. In this model, the half-skyrmion matter is  in the phase where the quark condensate $\la\bar{q}q\ra$ vanishes on the average in the unit shell, but locally non-zero, thereby producing an inhomogeneous structure.

The situation is different when the dilaton is present. Consider adding the soft dilaton to the Skyrme model of \cite{park-am} that accounts explicitly for spontaneously broken scale symmetry which gets restored at the chiral transition density $n_\chi$. We shall call this ``Dskyrmion" model. The Dskyrmion Lagrangian is constructed such that the scale symmetry is spontaneously broken by a Coleman-Weinberg-type potential $V(\chi)$. It can be obtained from the dHLS Lagrangian (\ref{dHLS}) by dropping the vector mesons and inserting the Skyrme term,
\begin{eqnarray}
{\cal L}_{sk}
&=& \frac{F_\pi^2}{4} \left(\frac{\chi}{f_\chi}\right)^2
{\rm Tr} (L_\mu L^\mu) + \frac{1}{32e^2}{\rm Tr} [L_\mu, L_\nu]^2\nonumber\\
&& +\frac{F_\pi^2}{4}\left(\frac{\chi}{f_\chi}\right)^3
{\rm Tr}{\cal M} (U+U^\dagger-2),
\nonumber \\
&&+\frac{1}{2}\partial_\mu \chi\partial^\mu \chi + V(\chi)\label{skyrme-lag}
\end{eqnarray}
where ${\cal M}$ is the quark-mass matrix, $V(\chi)$ is the potential that encodes the trace anomaly involving the
soft dilaton, $ L_{\mu} = U^{\dagger} \partial_{\mu} U $, with
$U$ the chiral field taking values in $su(2)$ and $f_\chi$ is the $vev$ of $\chi$.

The results of the analysis~\cite{LPR-halfskyrmion,pionvelocity} are as follows:
\begin{enumerate}
\item The skyrmion-half-skyrmion phase transition takes place at a density $n_{1/2}$ slightly above $n_0$. The precise value of $n_{1/2}$ does not depend sensitively on the dilaton mass for $m_\chi\lsim 1$ GeV. For a light dilaton mass appropriate for the soft component, i.e., $\sim (600-700)$ MeV, which is also reasonable for nuclear phenomenology in the vicinity of nuclear matter density, the transition takes place in the range $1.3 \lsim n_{1/2}/n_0\lsim 2$.
\item At $n_{1/2}$, the quark condensate $\la\bar{q}q\ra\propto {\Tr}U\rightarrow 0$ but the pion decay constant $F_\pi^*$ remains non-zero, dropping slowly at increasing density up to the chiral transition density $n_\chi$, signalling that the half-skyrmion phase is in the Nambu-Goldstone mode. The chiral restoration density $n_\chi$ is on the contrary highly sensitive to the value of the dilaton mass. For the mass adopted here, it comes out to be $n_\chi\gsim 4n_0$.
\item In the half-skyrmion phase, the nucleon mass scales with the dilaton condensate $\chi_0\equiv  \la\chi\ra$ -- which drops slowly, not with the quark condensate. This is similar to the behavior of the technibaryon  discussed recently as a source for dark matter~\cite{ellis}. It remains non-zero up to the critical density $n_\chi$. The dilaton mass scales also linearly in $\la\chi\ra$.

\end{enumerate}

The phase structure given by the Dskyrmion model is then as follows. For (a) $0 < n <n_{1/2}$, the system is in the skyrmion phase, i.e., in the standard Nambu-Goldstone mode with a non-vanishing quark condensate and pion decay constant; for (b) $n_{1/2} < n < n_\chi$, the system is in the half-skyrmion phase with the vanishing $\la\bar{q}q\ra$ but non-zero pion decay constant. Chiral symmetry is still broken with an order parameter not characterized by $\la\bar{q}q\ra$  but by a $vev$ of higher dimension operators; for (c) $n>n_\chi$, the system is in standard Wigner-Weyl phase with vanishing quark condensate and pion decay constant. Our model cannot access this phase.

This half-skyrmion structure at $n_{1/2}$ is expected to bring about substantial changes at densities above normal nuclear matter density in hadronic properties, in particular, the Brown-Rho (BR) scaling proposed in \cite{BR91}. In this paper we focus on nuclear tensor force in which scaling has a strong influence as first argued in \cite{BR-tensor}.

In order to write the new scaling (dubbed ``BLPR" for short), we need to bring in the vector degrees of freedom. The pertinent Lagrangian is the dHLS (\ref{dHLS}). Up to date no consistent and trustful calculations putting this Lagrangian on crystals are available~\footnote{A systematic work is in progress on this problem. There have been performed several calculations (involving one of the authors (MR)) with HLS Lagrangian with or without the dilaton scalar. Unfortunately, however, none of them is fully consistent with the symmetries involved: The homogeneous Wess-Zumino terms that are present in the presence of vector mesons are not correctly treated. We therefore do not rely on the results so obtained.}. We can however exploit that HLS is gauge-equivalent to the non-linear sigma model and deduce their scaling based on what is known of HLS.
HLS predicts that approaching the chiral transition at $n_\chi$, the vector meson mass scales as~\cite{HY:PR}\footnote{We have not verified that one can  approach the vector manifestation (VM) fixed point at which chiral symmetry is restored from the half-skyrmion phase. Hence the scaling (\ref{newBR-V}) has not been proven on crystal lattice. We are simply assuming it. One can however surmise what could happen. As stated above, the quark condensate $\la\bar{q}q\ra$ is zero in the half-skyrmion phase while with the dilaton field present, chiral symmetry is not necessarily restored. One way to interpret the scaling (\ref{newBR-V}) in the context of skyrmion crystals is that in the half-skyrmion phase, $\la\bar{q}q\ra\sim {\Tr}U={\Tr}U_1 +{\Tr}U_2=0$ when averaged over the cell with the two half-skyrmions giving ${\Tr}U_1=c$ and ${\Tr}U_2=-c$ with $c\neq 0$ and that as one approaches the VM fixed point, $c\rightarrow 0$. Possible non-vanishing chiral order parameters on the crystal -- conjectured above -- could carry this quantity $c$ in such a way that they vanish at $n_\chi$.}
\be
 m_\rho^*/m_\rho\approx g_\rho^*/g_\rho\approx \la\bar{q}q\ra^*/\la\bar{q}q\ra.\label{newBR-V}
\ee
This implies that the $\rho$ mass vanishes as the quark condensate goes to zero. If the local field description for meson degrees of freedom makes sense as one approaches chiral restoration, this would imply vanishing mass at the chiral phase transition density $n_\chi$.

If the flavor $U(2)$ is not a good symmetry in medium -- which is likely the case at a density $n\gsim n_{1/2}$ -- for the $\rho$ and $\omega$, the scaling of the $\omega$ mass need not be the same as that of the $\rho$. The vector manifestation may not apply to the singlet vector meson. However it seems plausible on the ground of ``mended symmetries" that the $\omega$ mass also goes to zero. How it does is not known.

It has been shown that (\ref{newBR-V}) holds in the presence of baryons whose masses are entirely dynamically generated~\cite{HKimR} (with the baryon mass going to zero at $n_\chi$). If on the other hand, the baryon mass remains massive as was found in the Dskyrmion model discussed above or the parity-doublet model with a large $m_0$~\cite{SLPR}, the baryons can be integrated out in approaching $n_\chi$, so we expect that the scaling (\ref{newBR-V}) should still hold.

The PLBR we shall study is the following.\footnote{We have replaced the dilaton $\chi$ by the scalar $s$ that represents the scalar that figures in nuclear physics. It will be mostly of $\chi$ at low density but could be different at high density with possible admixing to glueballs.}
\begin{itemize}
\item For the ``low density regime (regime-I)" $0<n<n_{1/2}$ , the scaling remains the same as in the old BR~\cite{BR91}:
\be
m_N^*/m_N&\approx& m_V^*/m_V\approx m_s^*/m_s\approx  F_\pi^*/F_\pi\equiv \Phi_I (n), \nonumber\\
g_V^*/g_V&\approx& 1\label{oldBR}
\ee
where $V$ stands for both $\rho$ and $\omega$.
\item For the ``high density regime (regime-II)" $n_{1/2}<n<n_\chi$, the scaling is changed to
\be
m_N^*/m_N&\approx& b,\nonumber\\
m_\rho^*/m_\rho&\approx& g_\rho^*/g_\rho \equiv \Phi_{II} (n),\nonumber\\
m_\omega^*/m_\omega&\approx& \Phi_\omega (n),\nonumber\\
m_s^*/m_s &\approx& \Phi_s (n).\label{newBR}
\ee
Since the nucleon mass is found to scale weakly, we set $b$ equal to a constant $b\lsim 1$. ??? There are four scaling functions $\Phi_{I,II,\omega,s}$ that can be different though they must be related to each other. The bdHLS model supplemented with the dilaton-limit and VM fixed points should provide  their behavior in density. For instance, although both the nucleon mass and the dilaton mass, at first sight, scale with the dilaton condensate $\la\chi\ra$, we can see from the analysis in \cite{sliding} that the scalar mass drops much faster than the dilaton condensate at density $> n_0$. There are several reasons to expect that $\Phi_s(n)<b$. First there seems to be a strong back-reaction from the soliton background on the fluctuation of the dilaton field whereas there can be a substantial-sized chiral-invariant mass $m_0$ in the nucleon mass as in the parity-doublet model~\cite{SLPR,PLRS} which prevents a rapid drop. More generally, locked to chiral symmetry, $m_s$ must join  the pion mass $m_\pi=0$ at $n_\chi$ in the chiral limit.

As for the scaling $\Phi_\omega$, if $U(2)$ symmetry held for $\rho$ and $\omega$, we would have $\Phi_\omega=\Phi_{II}$. We expect it not to hold in medium for $n\gsim n_{1/2}$ and hence we have no guidance from the vector manifestation, the only possibility that we have being that it vanishes at the chral restoration density..

Given that we have no reliable numerical results for the scaling functions, for numerical analyses, we shall simply choose reasonable values for $c_x$ in the form
\be
\Phi_x=\frac {1}{1+c_x\frac{n}{n_0}}\label{Phis}
\ee
for $x=I, II,s,\omega$.
\end{itemize}
\subsubsection{\it Tensor forces}
\indent\indent A simple prediction as to how tensor forces behave in nuclear and dense matter can be deduced from the above scaling relations~\cite{LPR-halfskyrmion}. Since the nucleon mass remains more or less unscaled, we can use non-relativistic approximation for the nucleon and write down the tensor forces contributed by the $\pi$ and $\rho$ exchanges. The well-known formulas are
\begin{eqnarray}
V_M^T(r)&&= S_M\frac{f_{NM}^2}{4\pi}m_M \tau_1 \cdot \tau_2 S_{12}\nonumber\\
&& \left(
 \left[ \frac{1}{(m_M r)^3} + \frac{1}{(m_M r)^2}
+ \frac{1}{3 m_Mr} \right] e^{-m_M r}\right),
\label{tenforce}
\end{eqnarray}
where $M=\pi, \rho$, $S_{\rho(\pi)}=+1(-1)$. Note that the tensor forces come with an opposite sign between the pion and $\rho$ tensors.

Following the common practice, we will leave the pion tensor unaffected by the density.\footnote{In fact it can be shown with the help of low-energy theorems that the pion tensor force remains little modified up to $n\sim (3-4)n_0$.} As for the $\rho$ tensor, apart from the scaling mass $m_\rho^*$, it is the scaling of $f_{N\rho}$ that is crucial. Written in terms of the parameters of the bdHLS Lagrangian (\ref{bdHLS}), we have the ratio
\be
R\equiv \frac{f_{N\rho}^*}{f_{N\rho}}\approx \frac{g_{VN}^*}{g_{VN}}\frac{m_\rho^*}{m_\rho}\frac{m_N}{m_N^*}
\ee
where we have defined the effective $V$-$N$ coupling $g_{VN}=g(1-g_V)$.
It follows from the scaling relations (\ref{oldBR}) and (\ref{newBR}) that
\be
R&\approx& 1\ \ \ {\rm for}\ \ \ 0\lsim n \lsim n_{1/2}\label{R1} \\
& \approx& \left(\frac{1-g_{V\rho}^*}{1-g_{V\rho}}\right){\Phi_{II}}^2 \ \ {\rm for} \ \  n_{1/2}\lsim n \lsim n_c.\label{R2}
\ee
We note that the ratio $R$ will be strongly suppressed for $n>n_{1/2}$. In addition to the suppression by the factor $\Phi_{II}^2$, the approach to the dilaton-limit fixed point would bring $(1-g_{V\rho}^*)$ go to zero. This makes the $\rho$ tensor killed rapidly.
\begin{figure}[ht!]
\begin{center}
\includegraphics[height=5.6cm]{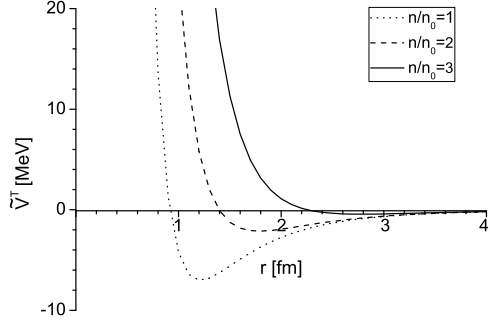}x
\includegraphics[height=5.6cm]{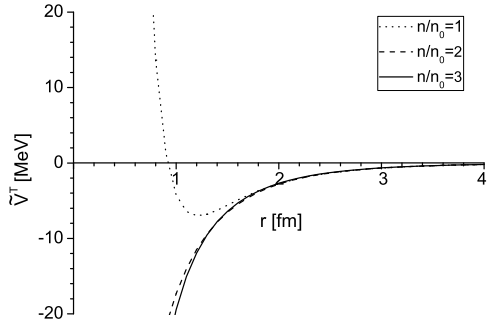}
\vskip -0.5cm
\caption{Sum of $\pi$ and $\rho$ tensor forces $\tilde{V}^T\equiv (\tau_1 \cdot \tau_2 S_{12})^{-1} (V_\pi^T +V_\rho^T)$ in units of MeV for densities $n/n_0$ =1, 2 and 3 with the ``old scaling" $\Phi \approx  1-0.15 n/n_0$ and $R\approx 1$ for all $n$ (left panel) and with the ``new scaling," $\Phi_I \approx 1-0.15 n/n_0$  with  $R\approx 1$ for $n<n_{1/2}$ and $\Phi_{II}\approx \Phi_I$ and $R\approx \Phi_{II}^2$ for $n>n_{1/2}$, assuming $ n_0<n_{1/2}<2n_0$ (right panel).}\label{tensor}
\end{center}
\end{figure}

What happens is illustrated in Fig.~\ref{tensor} both with the ``old BR" (no topology change) and the ``new BR" for BLPR (with topology change). With the old BR where the skyrmion-half-skyrmion phase change is not taken into account, the net tensor force decreases continuously in density with the attraction vanishing at $n\sim 3n_0$ for the given parameters. In a stark contrast, with the BLPR, the $\rho$ tensor force nearly disappears at $n\sim 3n_0$ leaving only the pion tensor active. Up to $n_{1/2}$, they behave the same way, so that the C14 dating beta decay\cite{c14} will be unaffected by the BLPR.
Once the density exceeds $n_{1/2}$, there is a drastic difference between the two scenarios. In the BLPR, the $\rho$ tensor gets rapidly suppressed and for a reasonable values of $c_x$, the pion tensor is the only component left. This will have an interesting consequence on the nuclear symmetry energy as mentioned below.
\subsubsection{\it Effect of BLPR on the symmetry energy}
\indent\indent If we assume as argued by several authors~\cite{BM,xu-li,polls} that the symmetry energy is dominated by the tensor force, then the main contribution to the symmetry energy factor, following Brown and Machleidt~\cite{BM}, can be written as
\be
S\sim \la V_{sym}\ra\approx \frac{12}{\bar{E}}\la V_T^2(r)\ra\label{BM}
\ee
where $\bar{E}\approx 200$ MeV is the average energy typical of the tensor force excitation and $V_T$ is the radial part -- with the scaling factor $R$ taken into account -- of the net tensor force. There are other contributions such as the kinetic energy term etc. They can be ignored for our purpose. We see immediately from the above discussion that the symmetry energy will decrease toward $n_{1/2}$ and then increase afterwards as the cancelation diminishes, giving a cusp at $n_{1/2}$. As described below, this cusp feature is supported by a pure skyrmion neutron matter when collective-quantized.

This prediction is in contrast with that of the old BR discussed in \cite{xu-li-BR}. In this article, the authors apply the old BR to explain the ``supersoft" symmetry energy which vanishes near $n\sim 3n_0$, namely, the $\alpha=+1$ curve in Fig.~\ref{a-1}. This comes about because the net tensor force decreases beyond $n\sim n_0$ in the old BR. In this paper we are arguing that the decrease is stopped by the presence of the half-skyrmion phase.  Experiments will decide which scenario is viable.
\subsubsection{\it Symmetry energy in the skyrmion-half-skyrmion phase transition}\label{sym-half}
\indent\indent In this subsection, we discuss the $\epsilon_{ym}$ obtained~\cite{LPR-halfskyrmion} by quantizing the skyrmion matter described by the Lagrangian (\ref{dHLS}). This confirms the correctness in transcribing the skyrmion crystal result into an effective Lagrangian.

In the skyrmion framework, the symmetry energy comes from a term subleading in $N_c$. It must therefore arise from the collective quantization of multi-skyrmion systems.  In his original work on the skyrmion crystal, Klebanov~\cite{klebanov} discussed how to collective-quantize the pure neutron system. We apply this method to the skyrmion matter as well as to the half-skyrmion matter we have obtained.

Consider an $A$-nucleon system for $B=A\rightarrow\infty$ (where $B$ is the winding number). Following Klebanov, the whole matter is rotated through a single set of collective
coordinates $U(\vec{r}, t) = A(t) U_0(\vec{r}) A^\dagger (t)$ where $U_0(\vec{r})$ is the static crystal configuration with the
lowest energy for a given density. The canonical quantization leads to
\be
E^{\mbox{tot}} = A M_{\mbox{cl}}
+ \frac{1}{2A \lambda_{I}} I^{\mbox{tot}} (I^{\mbox{tot}}+1),
\ee
where $M_{\mbox{cl}}$ and $\lambda_{I}$ are, respectively, the mass and the isospin moment of inertia

which is given as an integral over the single shell of a function consisting of the skyrmion configuration $U_0$.

Since $I^{\mbox{tot}}$ is the total isospin which would be the same as the third component of the isospin $I_3$ for pure neutron matter,  this suggests taking\footnote{{This is equivalent to assuming that $C_I$ in Clebsch-Gordan series, Eq.~(\ref{CGseries}), is dominated by $I = 1/2 |N_n -N_p|$.  It is definitely the case with $N_n >> N_p$ or $N_p >> N_n$.}}  for $\delta\equiv (N-P)/(N+P)\lsim 1$
\be
I^{\mbox{tot}}=\frac 12 A\delta.
\ee
Thus the energy per nucleon in an infinite matter ($A=\infty$) is
\be
E=E_0 +\frac{1}{8\lambda_I}\delta^2.\label{E}
\ee
with $E_0=M_{cl}$.
This leads to the symmetry energy factor
\be
S\approx \frac{1}{8\lambda_I}.\label{S-skyrmion}
\ee
Again the kinetic energy contribution (which is $1/N_c$-suppressed and expected to be unimportant at high density) is ignored in this expression. The numerical results for the given parameters are plotted for densities below and above $n_0$ in Fig.~\ref{skyrme-SE}. What is noticeable is the cusp at $n_{1/2}$, which is similar to what is expected with the BLPR as discussed above. This feature seems to be considerably smoothed out by nuclear many-body correlations in microscopic nuclear field theory approach with $V_{lowk}$ and old BR scaling incorporated into renormalization-group flow. It leaves, however, its distinctive imprint in the change in the symmetry energy from soft to hard at the density of topology change.
\begin{figure}[hbt]
\centerline{
\includegraphics[width=0.5\textwidth,angle=-90]{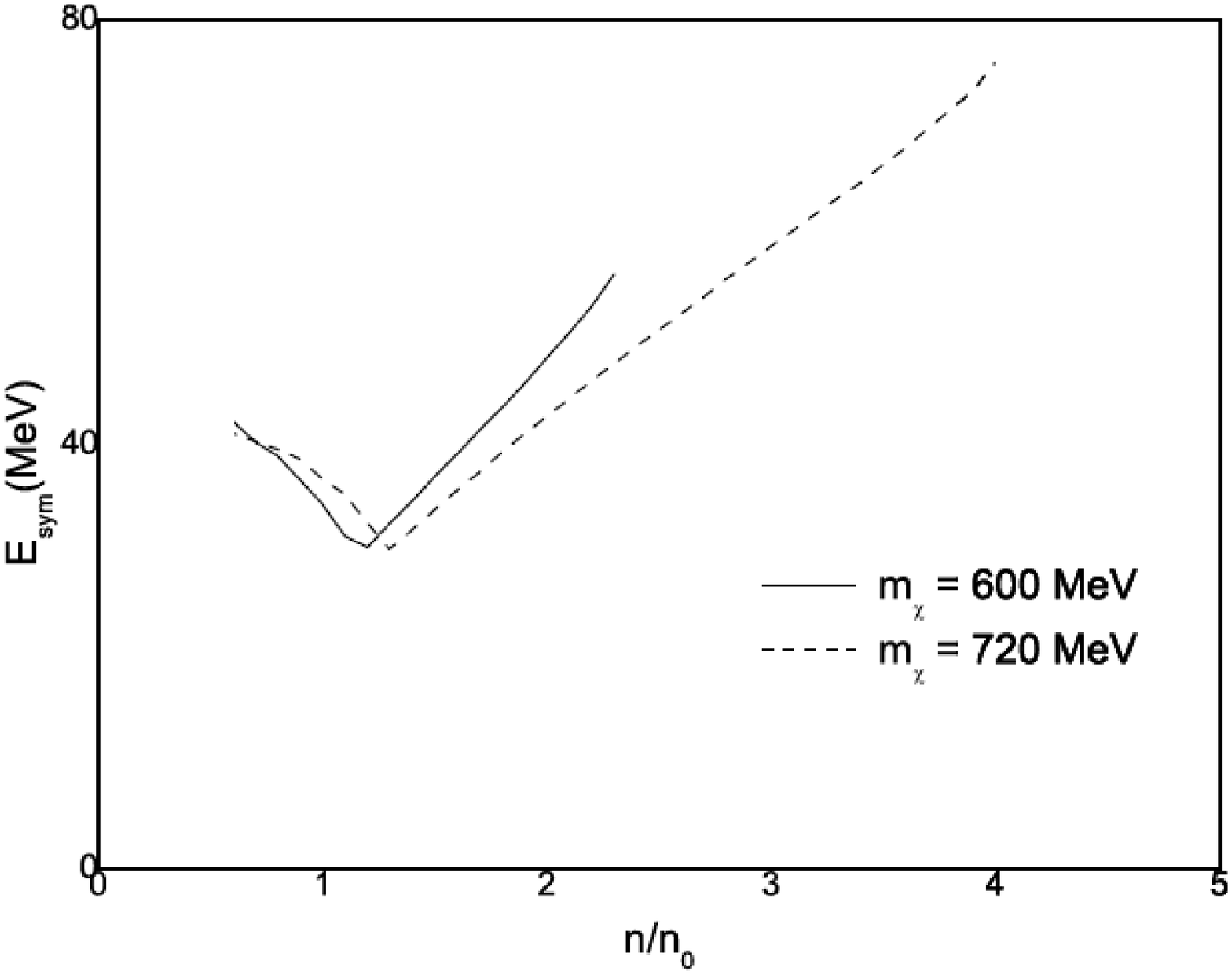}
}
\vskip -0.cm
\caption{Symmetry energy given by the collective rotation of the skyrmion matter with $f_\pi=93$ MeV, $1/e^2\approx 0.03$ and two values of dilaton mass. The cusp is located at $n_{1/2}$. The low density part that cannot be located precisely is not shown as the collective quantization method used is not applicable in that region.}
\label{skyrme-SE}
\end{figure}

A few caveats with this prediction are in order here.

First of all the Klebanov collective quantization method is not expected to be applicable for very low densities, so one cannot take seriously the results below the density $n_0$. Given that the Dskyrmion model (that is, the Skyrme model with dilaton) for the energy $\epsilon_0(n)$ at the classical level cannot describe nuclear matter saturation, one may ask how one can trust the calculation of $\epsilon_{sym}$. The possible answer to this question is that what we are calculating for $\epsilon_{sym}$ is a $1/N_c$ effect like the $N$-$\Delta$ mass difference since it is $\propto 1/\lambda_I$ and the moment of inertia $\lambda_I\propto N_c$. What enters into $\lambda_I$ is the leading $N_c$ term and the subleading effects that figure in the saturation of symmetric nuclear matter given by the energy $E_0$ would not affect $\epsilon_{sym}$ to the leading order we are considering. An evidence for this is the fact that the location of the density $n_{1/2}$ is extremely insensitive to the dilaton mass as one can see in Fig.~3 of \cite{LPR-halfskyrmion} whereas the $E_0$ -- hence the density $n_\chi$ at which $\la\bar{q}\ra^*=f_\pi^*=0$ -- is strongly affected by the dilaton mass.
\subsubsection{\it Dilaton-limit fixed point and mean-field theory}
\indent\indent The relativistic mean-field methods of the type mentioned in Sect.~\ref{RMF} assume that the parameters of the Lagrangian fixed in the matter-free vacuum remain unmodified when extrapolated to high density in the mean field. As mentioned above, doing mean field calculation with an effective Lagrangian endowed with pertinent symmetries for nuclear matter is justified because it is equivalent to Landau Fermi-liquid fixed point theory~\cite{friman-rho}. It is however not clear that it applies equally (1) to higher density departing far from the Fermi-liquid fixed point and to (2) the number of flavors greater than two involving strangeness in addition to up and down quark flavors.

Now in the standard RMF approach, possible modifications of the Lagrangian parameters due to the presence of medium are incorporated by the mean-field of higher-dimension operators. In the ``sliding-vacuum" adopted in this paper, such high-dimension effects are albeit approximately captured in the density dependence of the parameters of the effective Lagrangian with hidden local symmetry~\cite{friman-rho,song}. Suppose one calculates the symmetry energy using then the Lagrangian (\ref{bdHLS}) in the mean field. We get
\be
S/n=\frac{{g_{\rho N}^*}^2}{2{m_\rho^*}^2}\,+\cdots
\ee
where $g_{\rho N}\equiv g(1-g_{V\rho})$. Here the ellipsis contains other contributions such as the kinetic energy term that we may ignore at high density. We think that the effective $\rho$ mass $m_\rho^*$ encapsulates most, if not all, of corrections from higher-dimension operators in the standard RMF Lagrangian.

Now let us see what happens when the system approaches $n_\chi$. The $\rho$ mass goes like $\sim g_\rho^*$. Therefore $S$ will go like
\be
S/n \sim (1-g_{V\rho}^*)^2\rightarrow 0
\ee
due to the dilaton limit fixed point (\ref{decoupling}). This highly non-perturbative feature encoded in the DLFP is absent in the standard RMF approach or in finite-order chiral perturbation theory.

This of course does not mean that the standard RMF is wrong in the relevant density regime. It would remain valid until the DL sets in, but that may be at higher density than relevant for the process. This could be the case if the DL were to set in at about the same density as the vector manifestation. What the DLFP means is that the symmetry energy must diminish at some high density within the hadronic phase. What happens at the cross-over (or phase-transitioned) to a quark-gluon phase is unknown.  It stops being relevant in any case once the quark-gluon degrees of freedom set in.

\section{BLPR and Compact Stars}\label{newbr-star}
In this section we apply the formalism developed above to compact stars, addressing the issue of the maximum mass vs. radius of neutrons stars. This will be a concise summary of the work in progress reported in \cite{dongetal}, to which we refer for references.

We are interested in implementing the BLPR in an effective field theory (EFT) for nuclear matter. In doing this we adopt the ``double-decimation" strategy~\cite{BR:DD}.
\begin{enumerate}
\item The first decimation consists of obtaining via RG equation the $V_{lowk}$ in free space by decimating to the scale $\Lambda_{lowk}$ that describes nucleon-nucleon interactions up to lab momentum to $\sim 300$ MeV. It is best for our purpose to think of doing this in terms of a generalized HLS Lagrangian with parameters with the intrinsic density dependence given by the BLPR.
\item
The second decimation is to do nuclear many-body calculation with this $V_{lowK}$ to the Fermi-momentum scale $\Lambda_{fermi}$. There are a variety of ways of doing this step. They all amount essentially to doing Landau Fermi-liquid fixed point theory and arrive at nuclear matter at equilibrium density $n_0\sim 0.16$ fm$^3$. This is an EFT well-justified up to density near $n_0$. This step fixes the scaling properties of the parameters in the Lagrangian up to near $n=n_0$. We extend the same scaling up to $n_{1/2}$ since it is not too high above $n_0$.
\item  The last step in our application is to smoothly extrapolate with the formalism to high densities and calculate the EoS for compact stars. In doing this, one can adopt chiral perturbation strategy and include n-body forces -- if one wishes -- for $n>2$, suitably introducing scaling parameters for the n-body forces. In \cite{dongetal}, the topology change is incorporated in a smooth manner in terms of the changes in intrinsic scaling at $n_{1/2}$. It however ignores other degrees of freedom that might enter, such as strangeness that manifests in terms of kaon condensation and strange quarks etc.
\end{enumerate}

What came out of this exercise can be summarized as follows.

With the scaling properties constrained by nuclear matter properties in the region-I and heavy-ion data available up to $n\sim 4.5n_0$ in the region-II, a topology change at $1.5 <n/n_0< 2 $ can account for large-mass compact stars with no intervention of other degrees of freedom than nucleonic. A pure neutron-matter calculation gives $M_{max}\approx 2.4M_\odot$ and $R\approx 11$ km. The interior density of the star comes to $\sim 5n_0$. This gives a rough idea since compact stars are not pure neutron matter. A calculation that takes into account fully the relevant star conditions is in progress and will be reported in a later publication. Given that the density reaches the regime where kaon condensation can take place, possibly precipitated by the topology change discussed below, the appearance of strange quarks will also have to be addressed.

\section{ New Degrees of Freedom: Strangeness and Quarks}
\indent\indent So far we have been concerned with the symmetry energy in the two-flavor sector with the relevant baryonic degrees of freedom given by proton and neutron. In compact stars, apart from leptons, other hadronic degrees of freedom are expected to enter. In this section, we will consider how kaons and hyperons can figure in the symmetry energy relevant to compact stars. For this, we relate the chemical potentials of the participating degrees of freedom to the symmetry energy.
\subsection{Electron and muon threshold}
\indent\indent
We first consider the proton-neutron system without strangeness in chemical  equilibrium in weak interaction with electrons ($e$) and muons ($\mu$). Denote the proton and neutron number densities by $n_p$ and $n_n$, respectively. In terms of the chemical potentials
 \be
 \mu_n=\frac{\del\epsilon(n_p,n_n)}{\del n_n}|_{n_p}, \ \ \mu_p=\frac{\del\epsilon(n_p,n_n)}{\del n_p}|_{n_n},
\ee
we have from Eqs.~(\ref{epsn}) and (\ref{sn})
\be
\mu_n - \mu_p = 4(1 - 2 n_p/n)S(n)=\mu_e = \mu_{\mu}\equiv \mu.\label{betaeq}
\ee
The system must be charge neutral, so
\be
n_p =n_e + n_{\mu} = \frac{1}{3\pi^2}[\mu^3 + \Theta(\mu - m_{\mu})(\mu^2 - m_{\mu}^2)^{3/2}] \label{neutral}
\ee
From Eqs~(\ref{betaeq}) and (\ref{neutral}), we get
\be
\frac{1}{2}( 1 - \frac{\mu}{4 S(n)})n  =  \frac{1}{3\pi^2}[\mu^3 + \Theta(\mu - m_{\mu})(\mu^2 - m_{\mu}^2)^{3/2}]. \label{neutbet}
\ee
This relates the chemical potential $\mu$ to the symmetry energy factor $S$ for a given density. We will see how this chemical potential figures in the EoS for compact stars.

\subsection{Strangeness degree of freedom}
\indent\indent The first ``exotic" degree of freedom that we consider is the anti-kaon, in particular $K^-$, that carries the strangeness. The pion could also figure but it is not likely to enter significantly into the process. We will ignore it here. Although in free space, $K^-$ (that we will simply refer to as ``kaon" unless otherwise noted) is massive, its mass is expected to drop in dense medium because of strong attractive interactions with the nucleons in medium and when it becomes light enough, then the strangeness changing processes $n\leftrightarrow p+K^-$ and $e^-\leftrightarrow K^-+\nu_e$ can take place. When the kaon mass drops sufficiently low so that it crosses the increasing electron chemical potential $\mu_e$ at increasing density, kaons will condense and soften the EoS~\cite{KN,BKR-kaon,politzer-wise}. If these processes take place in chemical equilibrium with protons, neutrons, electrons and kaons in stellar matter, then
\be
\mu_n-\mu_p=\mu_e=\mu_K \equiv \mu.
\ee

There have been a large number of discussions on the possible impact of kaon condensation in compact stars and we will not enter into details of the theoretical tools employed. See for a review \cite{lattimeretal}. We will focus here on certain novel aspects of the process that takes place in the hadronic sector that we consider to be of relevance but have not been so far discussed. We will come to the quark-gluon sector later. We shall also postpone to later the role of strangeness that manifests through hyperons.

To illustrate the basic point, we sketch in the simplest form what is involved. For this we first consider chiral perturbation theory. Write {the grand potential, $\Omega$} of nuclear matter with s-wave kaons as
\be
\Omega(n,x, \mu, K^2)/V=\Omega(n,x,0)/V - a(n,x,\mu)K^2+\cdots
\ee
where we have expanded {the grand potential}
up to quadratic order in the {\em vev} of the kaon field, the ellipsis standing for higher orders in $K$.  To ${\cal O}(p^2)$ in the chiral expansion with baryon chiral Lagrangian, we can write
\be
a(n,x,\mu)&=&\mu^2 -m_K^2+\Pi(n,x,\mu),\label{quadraticterm}\\
 \Pi(n,x,\mu)&\equiv& \frac{1+x}{2f}\mu n +\frac{\Sigma_{KN}}{f^2}n\label{self-energy}
\ee
where $x=n_p/n$, $f\approx F_\pi$ and $\Sigma_{KN}$ is the $KN$ sigma term
\be
\Sigma_{KN}\approx  \frac 12 (\bar{m}+m_s)\la N|\bar{u}s+\bar{s}s|N\ra\ \label{sigmaterm}
\ee
where $\bar{m}=(m_u+m_d)/2$ with the subscripts $u$, $d$ and $s$ standing, respectively, for up quark, down quark and strange quark.
In chiral perturbation theory, the third term -- known as Weinberg-Tomozawa (WT) term -- corresponds to ${\cal O}(p)$ term and the sigma term to ${\cal O}(p^2)$. From the point of view of renormalization-group equation, what governs the flow to kaon condensation is the WT term with the sigma term being irrelevant.

Note that the kaon condensation threshold density $n^t_K$ for a given $\mu$ will be given by
\be
a(n=n^t_K,x,\mu)=0.
\ee

We should inject one remark here, returning to more discussions later. In the absence of other model-independent tools, chiral perturbation approach has been used to address the problem both for finite nuclei and infinite matter, in particular kaon condensation at high density. Unfortunately lacking lattice information, the reliability of chiral perturbation theory in medium, while reasonably successful for free-space processes, has not been established. Dense matter is a highly correlated system that is difficult to access by perturbative methods. For instance in kaon nuclear systems, while low-order chiral perturbation calculations, unitarized or otherwise, tend to give a shallow kaon-nuclear potential whereas phenomenology indicates an attraction several times typical chiral perturbation predictions~\cite{gal}. This deeply attractive potential in nuclear matter can be understood in a Walecka-type mean-field approach~\cite{meanfield}. Indeed the deeply bound kaon-light nuclear systems proposed by Akaishi and Yamazaki~\cite{yamazaki} require certain nonperturbative mechanisms not caught in low-order chiral perturbation expansion. In this connection, there is a hint from explorative lattice QCD simulation of kaon condensation~\cite{savage} on dense kaon systems containing up to 12 $K^-$'s which indicate that the properties of the condensate are remarkably well reproduced by leading-order chiral perturbation theory.  While direct baryonic background, difficult to implement in lattice calculations, is still needed, what this surprising result indicates, somewhat persuasively, is that there can be substantial cancelations in (numerous) higher chiral-order terms that cannot be captured by a few perturbatively calculable terms. This also questions the validity of ``higher-order" calculations in discussing the depth of kaon-nuclear potentials.

In the next subsection, we will present one possible non-perturbative effect connected with the half-skyrmion phase.

To proceed, let us simply assume that higher chiral-order and higher-K field terms can be ignored, and compute the proton and neutron chemical potentials modified by the kaon condensation
\be
\mu_n &=& \mu_n^0 - \left[ \frac{\mu}{2 f^2} + \frac{\Sigma_{KN}}{f^2} \right] K^2, \\
\mu_p &=& \mu_p^0 - \left[ \frac{\mu}{f^2} + \frac{\Sigma_{KN}}{f^2} \right] K^2,
\ee
where
\be
\mu_n^0 - \mu_p^0 = 4 (1 - 2\frac{n_p}{n_N}) S_N(n_N).
\ee
In beta-equilibrium with kaon condensation, the baryon number density is still carried by the nucleon, $n =n_B$, while strangeness is carried entirely by kaons.  If one ignores the back-reaction of kaon-nuclear interactions on nuclear interactions, then we can write the chemical potential difference using $S(\rho)$ and $K$,
\be
\mu_n-\mu_p = 4 (1 - 2x) S_N(n) + \frac{\mu}{2 f^2} K^2.\label{modification}
\ee

Equation (\ref{modification}) indicates that the symmetry energy will be modified when kaons condense. To see how it should be modified, we redefine the symmetry energy in the presence of kaon condensation as
\be
E(n,x, K^2) &=& E(n, 1/2,K^2) + \tilde{E}_{sym}(n, x, K^2)\,  \\
\tilde{E}_{sym}(n, x,K^2) &=& E(n,x,K^2) - E(n, 1/2,K^2).
\ee
Then the symmetry energy density  is given by
\be
\tilde{\epsilon}_{sym}(n, x,K^2) = n(1-2x)^2 S_N(n) + [\mu^2(n,x) - \mu^2(n,1/2)] K^2,
\ee
where the first term is the symmetry energy of pure neutron matter
\be
\epsilon^N_{sym}(n, x) =n (1-2x)^2 S_N(n)
\ee
and the chemical potential $\mu(n,x)$ is given by the solution of
\be
\mu^2 -m_K^2+ \frac{1+x}{2f}\mu n +\frac{\Sigma_{KN}}{f^2}n=0.
\ee
The numerical calculations  in a model adopted in this work, shown in Fig.~\ref{mux}, can be approximated as
\be
\mu(n,x) \sim a(1-bx) m_K,
\ee
where $a(n)$ and $b(n)$ (of order of $10^{-1}$ for $n/n_0 = 1- 6$)  are density-dependent  constants.

 \begin{figure}[t!]
\begin{center}
\includegraphics[height=7.0cm]{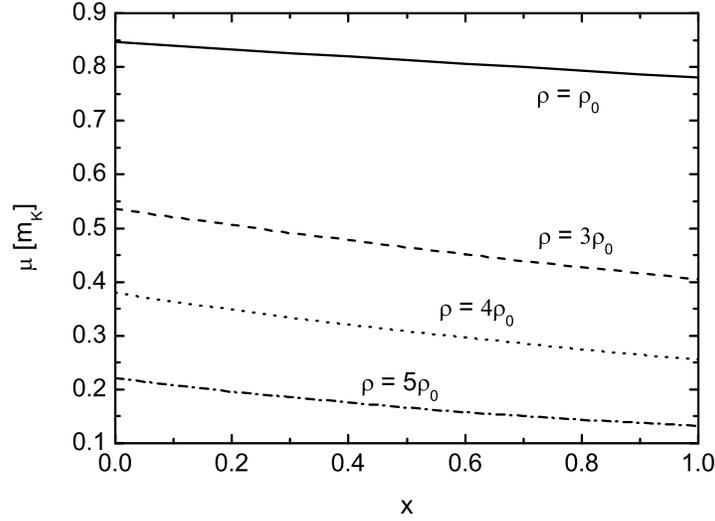}
\caption{$\mu(n,x) = \omega_K(x,\rho)$ with a parameter set used in \cite{KLR}.}
\label{mux}
\end{center}
\end{figure}

On the other hand the kaon chemical potential can be formally derived from the Lagrangian we are using  as a function of density given by
\be
\mu(n,x) &=& \frac{1}{2}[\frac{n_K}{K^2} -\frac{(1+x)n}{2f^2}]\label{munx} \\
&=& a^*(1 - b^* x)
\ee
where  $n_K$ is the kaon number density and
\be
a^* = \frac{1}{2}[\frac{n_K}{K^2} - \frac{n}{2f^2}], ~~ b^* =\frac{n}{2f^2} /[\frac{n_K}{K^2} - \frac{n}{2f^2}].
\ee
The numerical calculations confirm the formal expression and give $a^* \sim  a m_K$ and $b^* \sim  b$. Then we get
\be
\mu^2(n,x) - \mu^2(n,1/2) \sim (1-2x)\frac{a^2b}{2}m^2_K.
\ee
It  shows that the x-dependence is not in the form of $(1-2x)^2$ but linear in $(1-2x)$, which  breaks n-p permutation symmetry.  The basic reason is that when $K^-$ condenses, the isospin is spontaneously broken and the Weinberg-Tomozawa term in Eq.~(\ref{self-energy}) brings in isospin asymmetry.  Thus the $K^{-}$ condensation gives an additional contribution to the symmetry energy in the form
\be
\tilde{E}_{sym}(n, x,K^2) = (1-2x)^2 S_N(n) + (1-2x)S_K(n,K), \label{esymK}
\ee
where
\be
S_K(n,K) = \frac{a^2(n)b(n)}{2}\frac{K^2 m^2_K}{n}. \label{sk}
\ee
Note that for the parameters given in Fig.~\ref{mux},  $S_K(n,K)$ is of order of a few MeV.  Hence the  effect is not significant when compared to  $S_N$.

 Although we do not know exactly how symmetry energy  will be modified  in  kaon-condensed matter, one of the major effects of kaon condensation inferred from this simple example is that the $x-$ dependence of symmetry energy deviates from $(1-2x)^2$.  Whether it can be measured at heavy ion machine including FAIR, NICA or KoRIA (or ``RAON") is an interesting question,  if it is plausible that  the time scale for nuclear matter at the collider is long enough to realize the kaon condensed matter via weak interaction.}

\subsubsection{\it In-medium kaon mass in the skyrmion crystal}
\indent\indent
In this subsection, we turn to the skyrmion description and present a nontrivial consequence of the half-skyrmion phase on the in-medium mass $m_K^*(=\mu)$ given by
\be
m_K^*(n,x)=\sqrt{m_K^2-\Pi(n,x,m_K^*)}.
\ee
and on kaon condensation in dense matter. The self-energy $\Pi$ is given by (\ref{self-energy}) to ${\cal O}(p^2)$ in chiral perturbation theory, but as mentioned, it is unlikely that ChPT is reliable when kaons can condense. As an illustration of how nonperturbative effects can figure in the process, we examine how kaon spectrum is affected by the skyrmion-half-skyrmion transition at density $n_{1/2}$. We consider kaon fields fluctuating on top of a dense skyrmion background given by (\ref{skyrme-lag}) as in the Callan-Klebanov bound-state model for hyperons~\cite{callan-klebanov,scoccola}. At high density, the skyrmion background is expected to be reliably described by $SU(2)$ skyrmions put on crystal as in \cite{PKR}. We briefly describe what was found in that work and then discuss its ramification for the dropping kaon mass in compact-star medium .

The Lagrangian (\ref{skyrme-lag}) is extended to three flavors by supplementing with the Wess-Zumino term and taking the chiral $U$ field as
\begin{equation}
  U(\vec x,t) = \sqrt{U_K(\vec x,t)} U_0(\vec x) \sqrt{U_K(\vec x,t)},
\label{CKansatz}
\end{equation}
\begin{equation}
  U_K(\vec x,t) =
  e^{\frac{i}{\sqrt{2}f_{\pi}}
\left( \begin{array}{cc}
0 & K \\ K^\dagger &
0 \end{array} \right)},\
  U_0(\vec x)
    = \left( \begin{array}{cc} u_0(\vec x) & 0 \\ 0 & 1 \end{array} \right).
\end{equation}
Here $u_0\in SU(2)$ is the skyrmion background with winding number $B$.
When expanded in terms of the kaon field $K$, the Lagrangian we are concerned with is given by
\be
{\cal L}={\cal L}_{\rm SU(2)}+{\cal L}_{\rm K-SU(2)}
\ee
where
the kaonic part of the Lagrangian  (\ref{skyrme-lag}) in the background of the skyrmion background $u_0(x)$ and the dilaton $\chi_0$, ${\cal L}_{\rm K-SU(2)}$, is of the form
\be
{\cal L}_{\rm K-SU(2)}=(D_\mu K)^\dagger D^\mu K -K^\dagger(m_K^2 +{\cal V})K +z(K^\dagger D^\mu K B_\mu +{\rm h.c.})+\cdots
\ee
where $D_\mu K=(\del_\mu +V_\mu)K$ with $V_\mu=\frac 12(\del_\mu u_0 u_0^\dagger+\del_\mu u_0^\dagger u_0)$, ${\cal V}$ is a time-independent potential term, $B_\mu$ is the baryon current and $z$ is a known constant in front of the topological Wess-Zumino term. Explicitly
\begin{equation}
{\cal L}_{\rm K-SU(2)}
= \alpha ( \partial_\mu K^{\dagger} \partial^\mu K)
+ i \beta ( K^{\dagger} \dot K - \dot K^{\dagger} K )
- \gamma K^{\dagger} K+\cdots
\label{kaonlag}
\end{equation}
where
\begin{eqnarray}
\alpha &=& \frac{\kappa^2}{4}(u_0+u_0^\dagger+2),\nonumber
\\
\beta &=& \frac{N_c}{16f_{\pi}^2} B^0(u_0+u_0^\dagger+2),\nonumber
\\
\gamma &=& \frac{\kappa^3}{4}m_K^2(u_0+u_0^\dagger +2)\label{coeff}
\end{eqnarray}
where $\kappa=\chi_0/f_\chi$ and $B^0$ is the baryon charge density.

The Lagrangian (\ref{kaonlag}) describes the property of the kaon fluctuating in the background of the skyrmion $u_0$ and the dilaton $\chi_0$. We can consider the skyrmion with any baryon number $A$. For $A=B=1$, hyperons are obtained~\cite{callan-klebanov,scoccola} when  collective-quantized by the $SU(2)$ rotator coordinate $A(t)$
\be
U({\mathbf r},t)&=&A(t)U_0({\mathbf r})A^{-1} (t), \nonumber\\
K({\mathbf r},t)&=&A(t){\tilde U}({\mathbf r},t)\label{quantization}
\ee
The mass difference between the proton and the hyperons $\Delta m\gsim 175$ MeV can be accounted for by the binding of up to three kaons. The binding is predominantly given by the Wess-Zumino term with the hyperfine splittings obtained from collective-quantization. In terms of the $N_c$ counting, the Wess-Zumino term gives the vibrational correction of ${\cal O}(N_c^0)$ and the hyperfine splitting of ${\cal O}(N_c^{-1})$ to the baryon mass. In \cite{callan-klebanov}, the pure Skyrme model was studied with some success. With the introduction of the vector mesons $\rho$ and $\omega$, one obtains an even nicer fit to all hyperons in the baryon octet multiplet~\cite{scoccola}. It is likely that one would obtain an even better description with the infinite tower as argued above but this has not been confirmed yet.

Here we shall discuss applying the above strategy to many-nucleon systems as was done in \cite{PKR}. Details have not been worked out but it is clear that the procedure is straightforward, provided that the $u_0$ for multi-baryon system is known. We shall take $u_0$ studied in \cite{PKR}. Ideally it would be best within our framework to embed kaons into $A$-body systems with $A\rightarrow \infty$ described by $u_0$ simulated with the skyrmions put on crystal using the Lagrangian (\ref{dHLS}). However this has not been done yet. We will therefore consider the simpler Lagrangian (\ref{skyrme-lag}). What we are particularly interested in is the effect of the nonperturbative phenomenon associated with the change-over from a skyrmion matter to a half-skyrmion matter at $n_{1/2}$ on the properties of the fluctuating kaons on top of the given dense background.

\begin{figure}[h]  
\centerline{\epsfig{file=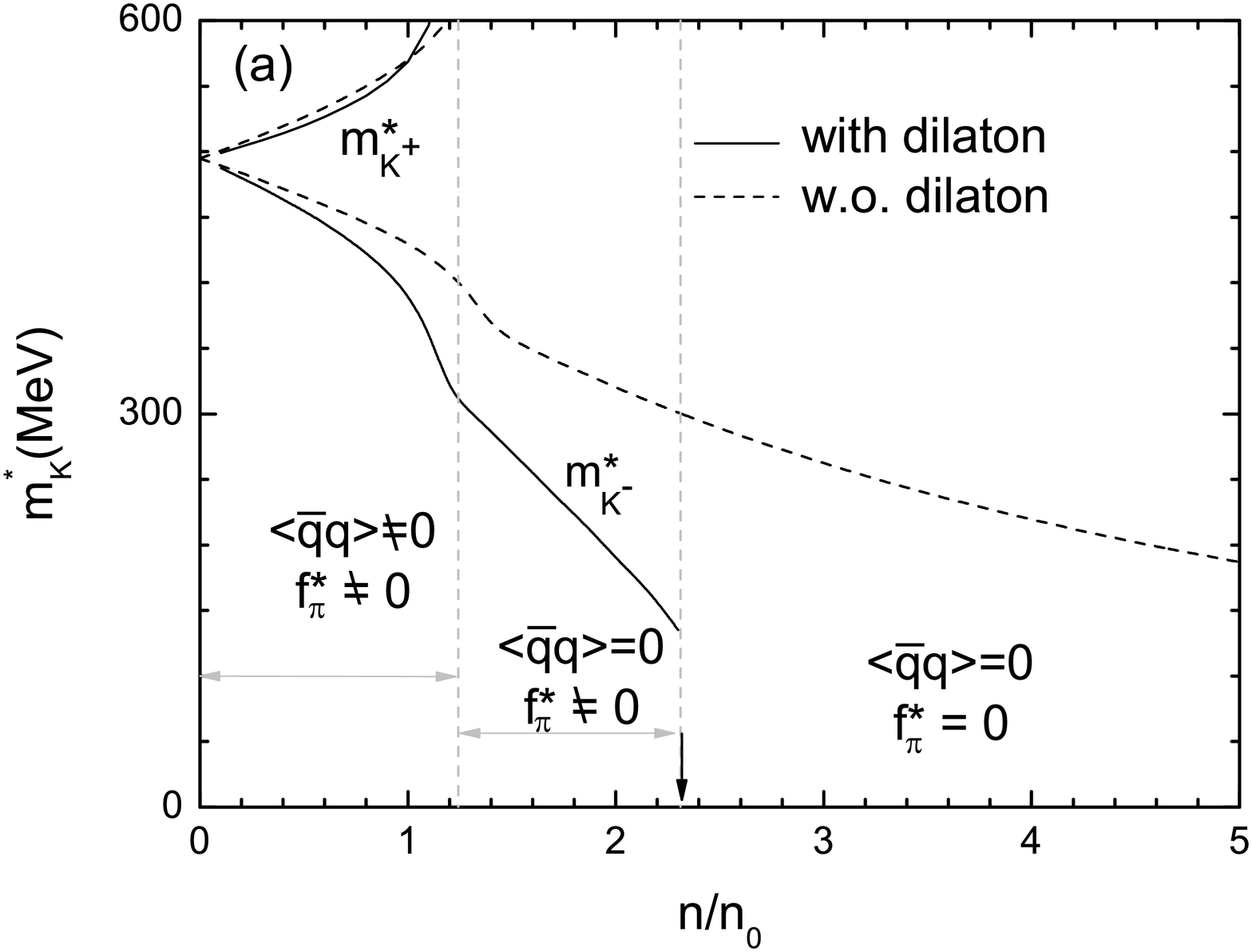, width=7.2cm, angle=0}
\epsfig{file=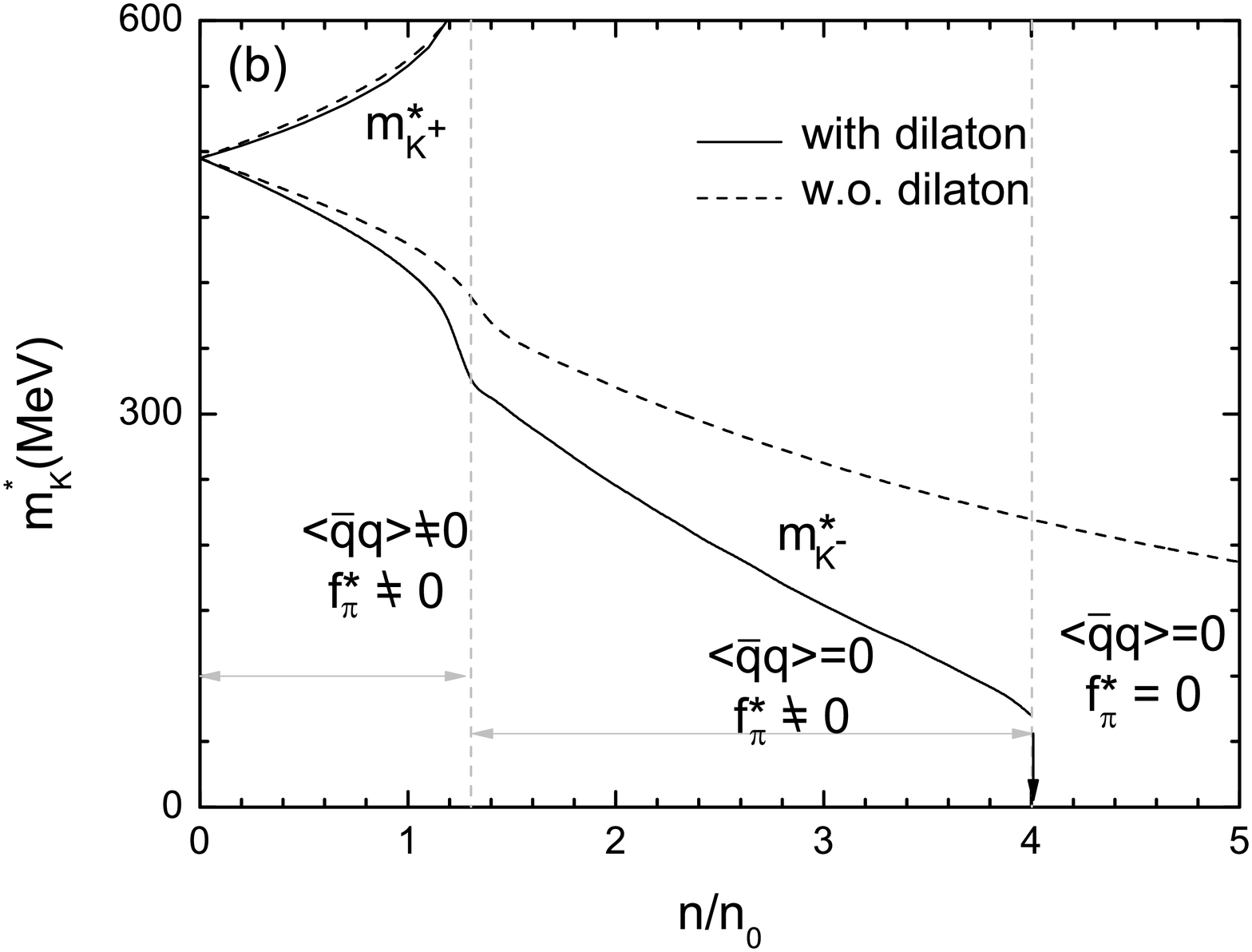, width=7.2cm, angle=0}}
\caption{ $m^*_{K^\pm}$ vs. $n/n_0$ (where $n_0\simeq 0.16$ fm$^{-3}$
is the normal nuclear matter density) in dense skyrmion matter
which consists of three phases: (I) $\la\bar{q}q\ra\neq 0$ and
$f_\pi^*\neq 0$, (II) $\la\bar{q}q\ra=0$ and $f_\pi^*\neq 0$ and
(III) $\la\bar{q}q\ra=0$ and $f_\pi^*=0$. The parameters are fixed
at $\sqrt{2}ef_\pi=m_\rho=780$ MeV and dilaton mass $m_\chi=600$ MeV
((a) left panel) and $m_\chi=720$ MeV ((b) right pannel). It should be noted that while $n_\chi$ is very sensitive to $m_\chi$, $n_{1/2}$ is practically independent of $m_\chi$. }
\label{fig1}
\end{figure}

To have a qualitative idea, one can simplify the calculation by averaging the quantities (\ref{coeff}) over the density-dependent skyrmion background as well as the dilaton condensate $\chi_0$. Then the s-wave kaon dispersion formula from (\ref{kaonlag}) is
\begin{equation}
\bar{\alpha} \omega_K^2 + 2 \bar{\beta} \omega_K + \bar{\gamma} = 0
\end{equation}
where $\bar{\alpha}$, $\bar{\beta}$ and $\bar{\gamma}$ are the averaged quantities. The solutions are given in Fig.~\ref{fig1} for two values of the dilaton mass with typical values for the parameters indicated in the figure caption.

One can see that the kaon mass drops similarly to the tree-order chiral perturbation results (\ref{self-energy}) up to density $n_0$, then undergoes a propitious drop not present in chiral perturbation theory at $n_{1/2}$ due to the appearance of the half-skyrmion phase.  Although the location of $n_{1/2}$ is quite insensitive to the dilaton mass as long as it is low $m_\chi\lsim 1$ GeV, the magnitude of the drop however does depend strongly on it. Such a drop -- if it is real -- would represent a genuinely nonperturbative effect brought in by the topology change\footnote{This is just about the same amount of a additional attraction required by Akaishi and Yamazaki~\cite{yamazaki} for their compressed nuclei.}.
\vskip 0.3cm

\subsubsection{\it Hyperons cannot appear before kaon condensation}\label{kconn}
\indent\indent  Kaons will condense in compact star matter when the electron  chemical potential $\mu_e$ is equal to the effective kaon energy or mass $m_K^*$ (for s-wave). As discussed above -- and one can see from Fig.~\ref{fig1}, kaon condensation in the skyrmion-crystal description will depend sensitively on the precise value of the dilaton mass. Without dilatons, $m_K^*$ will not fall low enough to condense up to density $n\sim 5n_0$, but with the dilaton of $m_\chi\sim 700$ MeV appropriate for nuclear physics~\cite{song}, $m_K^* \lsim 200$ Mev for $n\lsim 2.5n_0$. This suggests that kaon condensation at a reasonable density is likely to take place in this picture. It is intriguing that this property in the half-skyrmion phase is quite close to what one finds by fluctuating from the vector manifestation fixed point of hidden local symmetry theory as discussed in \cite{kaon-vm}.\footnote{The behavior of $m_K^*$ at $n>n_{1/2}$ can be compared with the tree-order chiral perturbation -- with BR scaling taken into account -- with $\Sigma_K\approx 250$ MeV, which is consistent with current lattice results (see \cite{KLR} for comments on this matter.). This would imply for electron chemical potential $\mu_e\approx 130$ MeV a critical density $n^c_K\sim 3.2n_0$, consistent with what was found in \cite{kaon-vm,meanfield}. A similar result was obtained by Westerberg~\cite{westerberg} using the Callan-Klebanov bound-state model embedded in a hypersphere. He found $n^c_K
 \sim 3.7n_0$.} Without the dilaton and the change-over from the skyrmion phase to the half-skyrmion phase, this would not occur. This again reminds one of the proximity of the half-skyrmion phase to the Georgi vector symmetry connected to the onset of the mended symmetry arrived at in the dilaton limit discussed above.

It should be emphasized that there has been an extensive debate with widely divergent conclusions as to what happens to kaon condensation when hyperons start figuring in the system. The issue is whether the role of the strangeness hyperons bring in is additional to that of condensed kaons and if so, their presence would not delay or even banish kaon condensation. The debate on this matter is typically based on mean-field treatments of phenomenological Lagrangians or low-order chiral Lagrangians supplemented with ``natural" higher-order meson field operators. But the problem is that such treatments cannot be trusted with confidence. That is because  such treatments suffer from one well-known defect: One does not know how to write an effective Lagrangian with higher-field operators that give in the mean-field a reliable description at higher density than that of normal matter. The customary procedure is to mix two unrelated ingredients, one anchored on low-order chiral perturbation theory and the other a phenomenologically implemented set of massive vector and scalar degrees of freedom. The two ingredients are unconstrained from each other, and hence although experiments guide theorists up to $n_0$, there is nothing that guides them beyond it, particularly at high density relevant to compact stars. This accounts for the widely varying predictions of EoS for $n>n_0$.

Here we would like to suggest that the Callan-Klebanov bound-state treatment used above offers a possible solution to this problem.

For this purpose, we can continue from Sec.~\ref{sym-half} where the symmetry energy was calculated by quantizing the neutron matter in a skyrmion-crystal description. We take $u_0$ to be a multi-skyrmion configuration for infinite matter put on crystal. We then embed kaons into that system and quantize the system using the collective coordinates (\ref{quantization}). In this way, binding $N_K$ kaons to the soliton with the winding number $A\gg 1$ would be tantamount to having hyperons in the system with the strangeness $N_K$. The binding comes mostly from the Wess-Zumino term which is the leading-order term involving strangeness, going as ${\cal O}(N_c^0)$ in the large $N_c$ limit. To this order, there is no distinction between the different hyperons for the given $N_K$ and there are no interactions between them. They will however appear at the next $1/N_c$ order as hyperfine effects accounting for splitting between members of equal strangeness.

In the CK scheme described above, it is easy to see that  hyperons cannot appear {\it before} kaons condense. To see this,  it suffices to first recall that kaons condense when the in-medium mass of the kaon $m_K^*$ is equal to the electron chemical potential $\mu_e$
 \be
 m_K^*=\mu_e\label{kacon1}
 \ee
and then note that in the CK model, hyperons appear when kaons get bound to the skyrmion, mainly by the Wess-Zumino term. The mass difference between the nucleon and the hyperon (denoted $Y$) in the CK model is
\be
m_Y-m_N \approx m_K^{eff} +{\cal O}(1/N_c)\label{hyp1}
\ee
where $m_K^{eff}$ is the effective mass of the kaon ($<m_K$) in the soliton background which is of ${\cal O}(N^0_c)$ in the $N_c$ counting. The hyperfine effect is in the ${\cal O}(1/N_c)$ terms. In compact stars, hyperons will appear when
\be
m_Y-m_N=\mu_e.
\ee
This is the same condition as kaon condensation (\ref{kacon1}) since in medium, $m_K^{eff}=m_K^*$.  Thus to ${\cal O}(N_c^0)$, hyperons appear at the density at which  kaons condense.  To go to higher order in $1/N_c$ in medium, one would have to consider  loop corrections that are density-dependent in addition to the hyperfine effects coming from the collective quantization. This would require a lot more work going beyond the present scheme, but it seems likely that the $1/N_c$ corrections will not upset this conclusion.

\subsubsection{\it Hyperons in standard RMF approaches}
\indent\indent As argued above, in the unified approach based on skyrmion crystal, the role of kaon condensation and that of hyperons are intricately related. In this case, it would be a double-counting if both were treated independently and combined. In the framework of phenomenological approach as well as in relativistic mean field approach, however, such a unified treatment of the two related mechanisms does not exist. What one does is to mix the two in an arbitrary way, unconstrained by each other.

Here we will simply ignore kaons and bring strangeness degrees of freedom only through hyperons. We may consider this as a formulation in terms of baryonic degrees of freedom only with the kaon degrees of freedom ``integrated out."

{For a system of nucleons and hyperons,   the hypercharge $Y=S+B$ (or strangeness number $S$) is conserved in strong interactions as are the proton $(I_3=1/2)$ and  neutron $(I_3=-1/2)$ numbers in $SU(2)$.  We can classify the eigenstate by the definite number of protons and neutrons and total number of hypercharge  $(I_3=1/2(N_p-N_n), Y = \Sigma_i Y_i N_{Y_i}$):
 \be
 |N_p, N_n, Y \ra.  \label{N}
 \ee
 Although  the $SU(3)$ symmetry is broken explicitly, the isospin symmetry can be used to decompose  the eigenstate, Eq.(\ref{N}), into the irreducible representations (multiplets) of $SU(2)$    as given by
 \be
 |N_p, N_n, \ra =  \Sigma_{(I,Y)} C_{(I,Y)} |I,Y: N_p, N_n\ra
\ee
where $ |\vec{I}|^2 = I(I+1)$. The energy of the eigenstate of the Hamiltonian does not depend on $I_3$ but on $I^2$ and $Y$ as  given by
\be
 E(N_p, N_n, Y)= \la N_p, N_n,Y | H |N_p, N_n,Y \ra  = \Sigma_{(I,Y)} |C_{(I,Y)}|^2  E_{(I,Y)}
\ee
where $E_{(I,Y)}$ is an reduced matrix element of the Hamiltonian, $H = H_0 + H_{int}$,
\be
E_{(I,Y)} = \la I,Y|| H ||I,Y\ra
\ee
which is independent of $I_3$, and depends on the details of the strong interactions for each $(I,Y)$-channel.

Therefore one can expect the nuclear symmetry energy can be affected by the presence of hyperons. The simplest parametrization inferred from the pure nuclear matter is
\be
E(n,x,y_i) = E(n, x=1/2, y_i) + \tilde{E}_{sym}(n_N, x, y_i), \label{symey}
\ee
where $y_i$ are the density of hyperon $i$ and $n_N$ is the nucleon number density $n_N = n_n + n_p$.   $n$ is the total baryon number density $n= n_N + y$, where   $y= \Sigma_i y_i= n - n_N$ is the total hyperon number density.

Let us consider first, as a simple example,  the   non-relativistic and non-interacting system.}   Although the kinetic contribution to the symmetry energy is small compared with the interaction term, particularly at high density,  one can still gain a valuable insight from it into how to parameterize the symmetry energy and how the presence of hyperons can affect it.

Recall that with only two flavors, n and p, the symmetry energy was defined in section 2.3 as
\be
E(n,x) &=& E(n, 1/2) + E_{sym}(n, x) \\
E_{sym}(n, x) &=& E(n,x) - E(n, 1/2),
\ee
where $E$ is the energy per baryon.  In the non-interacting Fermi-gas description,
\be
E^{free}_{sym}(n, x) &=& E_F[2^{2/3}\left((1-x)^{5/3} +x^{5/3}\right) -1] \\
E_F &=& \frac{3}{5} \frac{1}{2m}\left(\frac{3\pi^2 n}{2}\right)^{2/3} = \frac{3}{5} E_F^0(\frac{n}{n_0})^{2/3} \\
E_F^0 &=& \frac{1}{2m}\left(\frac{3\pi^2 n_0}{2}\right)^{2/3},
\ee
where $x= \frac{n_p}{n}$.  This can be approximated for the full range of $x$ from 0 to 1 in a simple form --within 5$\%$ accuracy -- as
\be
E^{free}_{sym}(n, x) &=& S^{free}(n)(1-2x)^2,\label{x2}\\
S^{free}(n) &=& (2^{2/3} -1)\frac{3}{5}E_F^0(\frac{n}{n_0})^{2/3}. \label{sfreeappp}
\ee
However when we take the above form with $(1-2x)^2$ as an expansion around $x=1/2$, then it cannot be naively used near $x\sim 0 $ (pure neutron matter) or $ 1$ for pure proton matter.  Nonetheless in the literature  the symmetry energy of the  above form has  been employed in many of parametrization schemes even for the pure neutron matter.  A recent discussion of this form for $x= 0-1$ can be found in \cite{CPR}\cite{dkm}.  We would like to see what comes out for the hyperonic matter in this simple free-gas approximation.   In fact,  we know very little of at what density the hyperonic matter can become relevant but let us simply assume that we have three types of hadrons, n, p and a hyperon (it can be any of the hyperons for the moment).   The $n$-$p$ symmetric state is defined by evenly distributed densities , $n_n=n_p $, in the presence of the hyperons with density $n_Y$. Then the energy per particle can be written as
\be
E(n, x, y) &=& E(n, x=1/2, y) + \tilde{E}_{sym}(n, x,y),\\
\tilde{E}_{sym}(n, x,y) &=& E(n, x, y) - E(n, x=1/2, y)\, .
\ee
Here $y$ is the fraction of hyperon number density, $y=n_Y/n$.  In  non-relativistic approximation,
\be
n E(n, x, y) &=& n_N \frac{1}{2} E_F^N [ (1-\alpha)^{5/3} + (1+\alpha)^{5/3}] + n_Y \frac{3}{5} \frac{1}{2m_Y}(3 \pi^2 n_Y)^{2/3}, \\
E(n, x, y)&=& \frac{n_N}{n} \frac{1}{2}\frac{3}{5}E_F^0(\frac{n_N}{n_0})^{2/3}[ (1-\alpha)^{5/3} + (1+\alpha)^{5/3}] + y^{5/3} \frac{3}{5} E_F^0 \frac{m}{m_Y} (\frac{2n}{n_0})^{2/3},\\
&=& [(1-y)^{5/3}  \frac{1}{2}[ (1-\alpha)^{5/3} + (1+\alpha)^{5/3}]  + 2^{2/3}\frac{m}{m_Y} y^{5/3}]\frac{3}{5} E_F^0 (\frac{n}{n_0})^{2/3} \\
\ee
with $\alpha = 1-2x$. Since $n_N = n - n_Y = (1-y)n$, the pure nuclear matter would correspond to $y=0$.   The symmetry energy from the kinetic energy part is
 \be
 \tilde{E}_{sym}(n, x,y) &=& E(n, x, y) - E(n, x=1/2, y) \\
&=& [(1-y)^{5/3}  \frac{1}{2}[ (1-\alpha)^{5/3} + (1+\alpha)^{5/3} -2]  ]\frac{3}{5} E_F^0 (\frac{n}{n_0})^{2/3}\\
&=& (1-y)^{5/3}S^{free}(n)(1-2x)^2,\label{sfreeappH}
 \ee
where $S^{free}(n)=(2^{2/3} -1)\frac{3}{5}E_F^0(\frac{n}{n_0})^{2/3}$.
Thus the effect of hyperons on the kinetic energy part of the symmetry energy is to reduce its strength by a factor of $(1-y)^{5/3}$. With nuclear interactions turned on, the contribution from the interactions will of course change the symmetry energy significantly.

It is convenient to parameterize the density dependence of $S$ in a polynomial form as
 \be
 S(n_N) = \Sigma_{\kappa} S_{\kappa} (\frac{n_N}{n_0})^{\kappa}, \label{plyn}
\ee
where $S_{\kappa}$ is a density-independent constant. In this form, the kinetic energy contribution corresponds to $\kappa = 2/3$,
 \be
 S_{2/3} = S^{free}_{kin} = (2^{2/3} -1)\frac{3}{5}E_F^0  .
 \ee
 To estimate the contributions from the nucleon-nucleon interactions, we assume that the interactions between nucleons are little affected by the presence of hyperons even beyond the hyperon threshold.  Then
\be
 \tilde{E}_{sym}(n, x,y) &=& E^N_{sym}(n_N)(\frac{n_N}{n}) = E^N_{sym}(n_N)(1-y)\label{1-y}\\
       &\equiv& (1-2x)^2 \tilde{S}(n,y).
\ee
This shows that the density in the argument of symmetry energy inferred from pure nucleon matter, $E^N_{sym}$,  should be the nucleon number density, not the total baryon number density and that the symmetry energy per baryon  becomes  smaller by a factor of $(1-y)$ since the baryon number is also carried by the hyperons. Now using
\be
 E^N_{sym}(n_N) = (1-2x)^2 S(n_N) = (1-2x)^2 \Sigma_{\kappa} S_{\kappa}(\frac{n}{n_0})^{\kappa} (1-y)^{(\kappa+1)}
\ee
we get
\be
 \tilde{S}(n,y) &=&  \Sigma_{\kappa} S_{\kappa}(\frac{n}{n_0})^{\kappa} (1-y)^{(\kappa +1)}.
 \ee
For the kinetic contribution, we get
\be
 \tilde{S}^{kin}(n,y) =  (1-y)^{5/3}S^{free}(n)
 \ee
 as discussed above.   What we learn from this simplified calculations is that
 \be
 \tilde{S}(n,y) \neq S(n).
 \ee

\subsection{Quark degrees of freedom}
\indent\indent
Similarly to the case of strange hadrons in dense hadronic matter, we might wonder whether one cannot expect more massive hadronic degree of freedom when density is much higher.  However, it appears that the next heavy hadrons, i.e.,  charmed hadrons, are too heavy to be excited at the density regime relevant to compact star interior.  Instead we expect a new form of matter in terms of quarks if the deconfinment phase transition can take place before other more massive hadronic degree of freedom appear.   The criterion  for the onset of the phase transition to a quark matter is whether the pressure of hadronic matter at a given density is greater than the pressure of quark matter. To address this issue, we would need reliable information on quark EoS.   The critical density will be determined at the phase boundary where the pressures are balanced. In our approach, we will suppose this phase change occurs at the end of kaon condensed nuclear matter.

The density at which quark matter presumably becomes relevant is much higher than the normal nuclear matter density, but it is not in the density regime where perturbative QCD can be applied. There can be a plethora of domains where hadronic and quarkish phases overlap, for which we have neither theoretical control nor experimental data. For our purpose, what is needed is a quarkish equivalent to the hadronic symmetry energy. As mentioned, the symmetry energy in the hadronic sector is controlled by strongly correlated interactions, with the kinetic energy term playing a minor role at high density. Whether similar correlations are involved in the quark sector is unknown. {For example, there have been no serious  discussions on the density dependence of the Clebsch-Gordon coefficients, $C_I$, for a given number of u, d and s quarks.  In the  SQM, it is assumed that $I=0$ is dominated.   The  quantities related to quark-quark correlations such as $ \Delta_{dd}$ or   $\Delta_{ud}$  analogous to  $ \Delta_{nn}$ or   $\Delta_{np}$ has not been sufficiently explored except for the case of the free quark-gas approximation.}

We will therefore explore a scenario of the phase change which bypasses the details of the EoS of quark matter {\em at the phase boundary with the assumed kaon condensed nuclear matter.}
\subsubsection{\it Non-strange quark matter}
\indent\indent
Quark matter can appear as a result of confinement-deconfinement phase transition.  This phase transition is basically a strong interaction process.  Suppose nuclear matter with no strangeness changes over to a deconfined quark phase. There will be no room for strange quarks in this case because it is the strong interaction that is involved. {At a later stage, it can evolve into a strange quark matter via weak transition emitting neutrons out of the system if SQM is more stable energetically for a given quark number density.}

We first consider nuclear matter without kaon condensation (NM for short) and deconfined non-strange quark matter (NSQM for short). Suppose there is a phase boundary between NM and NSQM at the density $n_c$ (for NM) and $n^Q_c$ (for QM)\footnote{We are taking into consideration that there may be a density jump between the NM and the QM at the boundary, such that $n^Q_c\neq 3n_c$.}, respectively, in the interior of a neutron star. The chemical equilibrium reads
\be
\mu_n = 2\mu_d + \mu_u, ~~~ \mu_p =  \mu_d + 2\mu_u. \label{chemeq}
\ee
The symmetry energies in each phase are related to the chemical potentials as
\be
\mu_n-\mu_p &=& 4 (1-2x_N) S_N(n_N)\\
\mu_d-\mu_u &=& 4 (1-2x_Q) S_Q(n_Q),
\ee
where $x_N = n_p/n_N$, $x_Q = n_u/n_Q$ and $n_Q = n_u + n_d$.
From the chemical equilibrium
\be
\mu_n -\mu_p &=& \mu_d-\mu_u,
\ee
using $x_Q = n_u/n_Q = (1+x_N)/3$, we get
\be
(1-2x_N) S_N(n_c) = \frac{1}{3}(1-2x_N) S_Q(n^Q_c) \label{xn}
\ee
from which we obtain a constraint on the symmetry energy at the chiral restoration density $n_c$ (which we assume to coincide with the deconfinement density),
\be
S_N(n_c)&=& \frac{1}{3}S_Q(n^Q_c)
\ee
for $x_N \neq 1/2$.
Just for an illustration, suppose we take an expression for the symmetry energy factor of the non-interacting quark gas:
\be
S_Q = \frac{3}{4}(3 \pi^2 n_Q /2 )^{1/3}[2^{1/3} -1],
\ee
and assume that the critical quark number density, $n^Q_c / 3$, is equal to the critical baryon number density, $n_c$,
then we will get
\be
S_N(n_c) = 24.6 (\frac{n_c}{n_0})^{1/3} \MeV\ . \label{snnc}
\ee
We may obtain a different constraint on the possible form of the nuclear symmetry energy, if nuclear matter undergoes a phase transition into a quark phase at $\rho_c$. However, that will yield a highly nontrivial constraint\cite{toro}. It also  depends on the nontrivial quark-quark interactions in the deconfined phase.
In a recent paper by G. Pagliara and J. Schaffner-Bielich \cite{PSB}, it is argued that the $S_Q$ can be three times larger than that given by the free quark model if the quark phase is in 2SC, which is of course a highly correlated system, unlike free quarks.   In Eq.~(\ref{xn}), the simplest solution is $x_N =1/2$ irrespective of the specific form of symmetry energy. We think this possibility can be realized in stellar matter at high density. We shall obtain this result from a different consideration.  However in heavy ion collision, $x$ is fixed to be  less than $1/2$, under which the phase transition occurs. So we have to solve nontrivial phase boundary conditions\cite{toro}\cite{CPM}\cite{PSB}.

\subsubsection{\it Strange quark matter}
\indent\indent  In dense compact-star matter, the system evolves in weak equilibrium, so strange quarks figure in dense matter through weak interactions.  The nature of interactions for the phase change is, however, different if it is from a phase in which kaons induced by the weak interaction are present to start with. Now if kaon condensation has taken place in the system, then there is already non-vanishing strangeness in the nuclear matter up to the phase boundary. In this case,  we can imagine the confinement-deconfinement phase transition taking place constrained by the weak equilibrium in the kaon condensed matter, leading to strange quark matter. In this subsection, we develop the scenario in which kaon condensation joins smoothly to a strange quark matter.

Consider the case where strange quark matter (SQM for short) meets kaon-condensed nuclear matter (KNM for short). As was done in the previous section, we can obtain the kaon threshold density, $n^t_K$, by the profile of the kaon chemical potential. From $n^t_K$ up to the critical density for chiral restoration $n_c$, the weak equilibrium condition reads
\be
\mu=4(1 - 2 x)S_N(n_c) + F(K,\mu,n_c).  \label{betaeq3}
\ee
{With the Weinberg-Tomozawa term for the kaon-nucleon interactions introduced in Section 4.2,  $F(K,\mu,n_c)$ can be expressed as}
\be
F(K,\mu,n_c) = \mu \tilde{F}(K,n_c),
\ee
and we have
\be
\mu=4(1 - 2 x)S_N(n_c) + \mu \tilde{F}(K,n_c).  \label{betaeq2}
\ee
As the kaon chemical potential -- equivalently effective mass -- $\mu$ approaches 0 at the critical density,  the solution $x=\frac{n_p}{n}=1/2$  appears naturally at the phase boundary,  provided $\tilde{F}(K,n_c) \neq 1$ which is assumed to be valid for the range of density we are concerned with.

At the phase boundary, the chemical equilibrium (via confinement-deconfinement) reads
\be
\mu_n -\mu_p &=& \mu_d-\mu_u,\label{chemq} \\
\mu_{K^-} &=& \mu_s - \mu_u\, . \label{chemk}
\ee
Note that the strange quark is required at the boundary, which implies that there should be a strange quark matter for $n > n_c$. Since $\mu(=\mu_K) =0$ , we have from Eqs.~(\ref{chemq}) and (\ref{chemk})
\be
\mu_u ~ = ~ \mu_d \ \ {\rm and} \ \ \mu_s ~ = ~ \mu_u,
\ee
which gives
\be
\mu_u = \mu_d = \mu_s. \label{musqm}
\ee
This is the chemical potential relation for the SQM in the massless limit. In this simple picture, the kaon condensed nuclear matter leads naturally to a strange quark matter.\footnote{The relation (\ref{musqm}) will of course be spoiled in nature with the s-quark mass. But going smoothly over from a kaon-condensed state to strange quark matter should remain intact.} In this sense, it is natural to consider the kaon condensed state as the {\em doorway} to strange quark matter as proposed elsewhere. The coincidence of kaon condensation and chiral symmetry restoration in the chiral limit was predicted a long time ago with a kaon bound to a skyrmion put on hypersphere~\cite{hypersphere}.
\section{Discussions and Conclusion}
\indent\indent
In this article reviewing our work done in the WCUIII/Hanyang program,  we discussed how  flavor symmetries figure in the nuclear symmetry energy which plays an important role in the structure of neutron-rich nuclei and in the high-density EoS of compact stars . In this way of looking at the problem, the prominent role of the tensor forces in nuclear interactions is highlighted. When dense baryonic matter is described by multi-skyrmions put on crystal, the structure of the tensor forces is controlled by a change in the parameters of our generalized hidden local symmetry Lagrangian which is engendered by a change of topology from the skyrmion matter to a half-skyrmion matter at a density $n_{1/2}$ lying slightly above the nuclear matter density. We suggest that this density regime could be explored by such RIB accelerators as the  ``KoRIA" (alias ``RAON")  being built in Korea.

The new scaling of the parameters, called ``BLPR scaling" replacing the old BR scaling, if correct, will be of crucial importance in the effort to access the regime of density where the chiral phase transition or deconfinement of quarks is expected to take place. We have found that the new scaling makes the EoS -- which is soft below $n_{1/2}$ -- become stiff above, precisely the mechanism needed to understand the 1.97$M_\odot$ star without invoking other degrees of freedom than nucleonic. On the other hand, the topology change at $n_{1/2}$ makes the effective anti-kaon (e.g., $K^-$) mass in medium to undergo a propitious drop so that it could speed up kaon condensation, which would then soften the EoS. We have also argued that hyperons can appear only when kaons condense, so they need not be considered as two different phenomena~\cite{LR-hyperons-kcond}. Since the stiffening of the nucleonic EoS defers kaon condensation to higher density, it is not clear what would be the net consequence between these two opposing effects. What we have found however is that even if kaon condensation takes place, it can play the role of a ``doorway state" to strange quark matter and compact star masses of $\sim 2 M_\odot$ can be accommodated.

We close with some additional remarks. 

One of the observations made with the skyrmion-half-skyrmion transition is the  cusp in the symmetry energy at $n_{1/2}$. The same cusp structure is expected when the crystal result is transcribed into a many-body theory language in terms of the BLPR and one assumes that the symmetry energy is dominated by the tensor forces. Such a cusp was not observed in standard phenomenological or relativistic mean field treatments. It is also possible that such a cusp structure will be smeared out by many-body correlations in microscopic many-body or field-theoretic approaches. Such a renormalization-group-implemented field theoretic calculation using $V_{lowk}$ with the BLPR suitably incorporated, discussed in \cite{dongetal}, does indicate the topology change taking place at $n_{1/}$ is smoothed by correlations.

Another observation that merits to be highlighted is that most of the parametric forms~\footnote{The exception is the ``supersoft" symmetry energy~\cite{supersoft} in which the repulsion changes into attraction making $\Delta_{nn}$ negative at $n\gsim 3n_0$. This feature may be explained by the old BR~\cite{xu-li-BR} but is at odds with the BLPR of this paper.} we have looked at are found to show $\Delta_{np}$, which is negative at normal nuclear density,  becoming positive at higher density, say, $\sim 3n_0$. This implies that the attraction between neutron-proton at lower density is changing into repulsion at higher density whereas the repulsion between neutron-neutron ($\Delta_{nn}$) keeps growing.

An intriguing consequence of the suppression of the $\rho$ tensor force at density $n\gsim n_{1/2}$  predicted in this paper is that at high density, $\gsim 3n_0$, the pion tensor will dominate. This could lead to a pion-condensed crystalline spiral structure predicted a long-time ago by Pandharipande and Smith~\cite{pandha}. Given that this structure arises due to the fractioning of skyrmions into half-skyrmions, this may be related to the ``baryonic popcorn" structure discussed in holographic approach to baryons~\cite{popcorn}.

Among the characteristics of the hidden local symmetry approach, a distinctive feature that is yet to be confirmed by either theory or experiments is the vector manifestation which predicts that at the chiral transition (in the chiral limit), a set of vector mesons become massless with a flavor gauge symmetry  ``emerging" from collective phenomena in dense and/or hot medium~\cite{HY:PR}. Although there are indirect evidences for it in the form of ``dropping mass," there is no direct or ``smoking-gun" evidence either for or against it. The highly heralded dilepton production in heavy-ion collisions is not a simple and trustful probe for in-medium vector-meson mass -- and hence chiral symmetry -- contrary to what has been  naively thought~\cite{nosmoking}. It remains to find an experiment that will do the job.

\section*{Acknowledgments}
We are grateful for helpful discussions with Masayasu Harada, Kyungmin Kim, Tom Kuo,  Won-Gi Paeng,  Byung-Yoon Park and Chihiro Sasaki. This work is supported by WCU (World Class University) program: Hadronic Matter under Extreme Conditions through the National Research Foundation of Korea funded by the Ministry of Education, Science and Technology (R33-2008-000-10087-0).  This work is partially based on the discussions during the YIPQS-WCU(Hanyang) joint international workshop on ``Dense strange nuclei and compressed baryonic matter" held at YITP in Kyoto, April 18-May 18 and also during the Second APCTP-WCU Focus Program ``From dense matter to compact stars in QCD and hQCD" at APCTP in Pohang, Aug. 16-25, 2011.

\end{document}